\documentclass[twoside,11pt]{article}

\usepackage{obs_study_style}

\usepackage{lscape}
\usepackage{float}
\usepackage{makecell}
\usepackage{pgf,tikz}
\usepackage{ulem}

\usepackage{csquotes}
\usepackage{hyperref}

\usepackage{colortbl}

\usepackage{multirow}
\usepackage{subcaption}

\usepackage{bbm}	
\usepackage{amsmath}
\usepackage{cleveref}

\usepackage{booktabs}

\newtheorem{assump}{Assumption}

\newtheorem{rem}{Remark}

\def\beq{\begin{equation}} 
\def\eeq{\end{equation}}
\def\beqn{\begin{eqnarray*}}
\def\eeqn{\end{eqnarray*}}
\def\Bitem{\begin{itemize}\setlength{\itemsep}{.2in}}
\def\bitem{\begin{itemize}\setlength{\itemsep}{.05in}}
\def\eitem{\end{itemize}}
\def\Benum{\begin{enumerate}\setlength{\itemsep}{.2in}}
\def\benum{\begin{enumerate}\setlength{\itemsep}{.05in}}
\def\eenum{\end{enumerate}}
\def\bmult{\begin{multline*}}
\def\emult{\end{multline*}}
\def\bcenter{\begin{center}}
\def\ecenter{\end{center}}
\def\bframe{\begin{frame}}
\def\eframe{\end{frame}}

\def\cO{\mathcal{O}}
\def\cP{\mathcal{P}}
\def\cQ{\mathcal{Q}}

\newcommand{\bbeta}{{\boldsymbol\beta}}

\newcommand{\btheta}{{\boldsymbol\theta}}

\newcommand{\E}{\operatorname{\mathbb{E}}}
\renewcommand{\P}{\operatorname{\mathbb{P}}}

\def\eps{\varepsilon}

\def\1{\mathbbm{1}}

\def\SS{\text{SS}}
\def\NS{\text{NS}}
\def\NN{\text{NN}}
\def\SN{\text{SN}}

\def\boldV{V}

\def\balpha{\alpha}
\def\bbeta{\beta}

\def\btheta{\theta}

\usetikzlibrary{arrows,shapes.arrows,shapes.geometric,shapes.multipart, decorations.pathmorphing,positioning,shapes.swigs,}

\definecolor{purple}{rgb}{0.4,.1,.9}

\DeclareMathOperator{\expit}{expit}

\newcommand{\eqn}{\begin{eqnarray}}
\newcommand{\ee}{\end{eqnarray}}
\newcommand{\eqnn}{\begin{eqnarray*}}
\newcommand{\een}{\end{eqnarray*}}

\heading{}{}{}{}{}{Andrew Ying, Ronghui Xu, Christina D. Chambers and Kenneth Lyons Jones}

\ShortHeadings{Causal Effects of Prenatal Drug Exposure on Birth Defects with Missing by Terathanasia}{Ying, Xu, Chambers and Jones}
\firstpageno{1}

\begin{document}

\title{Causal Effects of Prenatal Drug Exposure on Birth Defects with Missing by Terathanasia}

\author{\name Andrew Ying \email anying@wharton.upenn.edu \\
       \addr Department of Statistics and Data Science\\
       The Wharton School\\
       University of Pennsylvania\\
       Philadelphia, PA 19104, USA
       \AND
       \name Ronghui Xu \email rxu@ucsd.edu \\
       \addr Herbert Wertheim School of Public Health and Department of Mathematics\\
       University of California San Diego \\
       La Jolla, CA 92093, USA
       \AND
       \name Christina D. Chambers \email chchambers@health.ucsd.edu\\
    \addr Department of Pediatrics and Herbert Wertheim School of Public Health\\
       University of California San Diego \\
       La Jolla, CA 92093, USA
    \AND
    \name Kenneth Lyons Jones \email klyons@ucsd.edu\\
    \addr Department of Pediatrics\\
     University of California San Diego \\
       La Jolla, CA 92093, USA
       }

\maketitle

\begin{abstract}
A recent cohort study revealed a positive correlate between major structural birth defects in infants and a certain medication taken by the pregnant women. To draw valid causal inference, an outstanding problem to overcome was the missing birth defect outcomes among pregnancy losses resulting from spontaneous abortion. This led to missing not at random since, according to the theory of ``terathanasia'', that a defected fetus is more likely to be spontaneously aborted. Other complications in the data included left truncation, right censoring, observational nature, and rare events. In addition, the previous analysis stratified on live birth against spontaneous abortion, which was itself a post-exposure variable and hence did not lead to causal interpretation of the stratified results. In this paper we aim to estimate and provide inference for the causal parameters of scientific interest, including the principal effects, making use of the missing data mechanism informed by ``terathanasia''. The rare events with missing outcomes led to multiple sensitivity analyses where the causal parameters can be estimated with better confidence in each setting. Our findings should shed light on how studies on causal effects of medication or other exposures during pregnancy may be analyzed using state-of-the-art methodologies.
\end{abstract}


\section{Introduction}

Our work was motivated by a recent observational study carried out by the North American Organization of Teratology Information Specialists (OTIS), on the use of a specific medication during pregnancy. Analysis of the data found that the proportion of liveborn infants with major birth defects was higher for women exposed to the medication compared to unexposed women with the same underlying diseases. 

The biological mechanism, if any, that could explain the elevated birth defect rates was not understood, due to the lack of a pattern of major birth defects as traditionally seen with other teratogens such as thalidomide. One possible explanation for the increased rate of birth defects in live born infants could be related to the concept of ``terathanasia''.  This refers to the natural screening process whereby malformed conceptuses are preferentially spontaneously lost in the early weeks of pregnancy, thereby reducing the birth prevalence of congenital malformations identified in live born infants. Expanding on this concept, \cite{warkany1978terathansaia} further suggested that a maternal exposure early in pregnancy could influence the process of terathanasia. He cited an animal study in which prenatal exposure to the medication thyroxine led to differential mortality of embryos with cleft lip, thus reducing the number of surviving offspring with cleft lip.  It is also possible that maternal exposure to an agent early in pregnancy could suppress the natural screening process of terathanasia by ``rescuing" or preventing spontaneous loss of malformed embryos. 
Recently \cite{santiago:etal} discovered that endometrial preparation with etanercept increased embryo implantation and live birth rates in women suffering from recurrent implantation failure during IVF. 
In this latter scenario, a lower rate of spontaneous abortion following exposure to a specific medication might lead to a lower number of malformed embryos lost, and subsequently would result in an increased number of congenital anomalies in those embryos that survived to live birth. 

Figure \ref{fig:dag} depicts the causal relationship and the temporal order of the relevant variables for the description above. Solid arrows and dashed arrows refer to definite and possible causal relationships, respectively. The dashed arrow from drug exposure to birth defects reflects the causal effect that we are interested in inferring. The solid arrow from birth defects to spontaneous abortion represents the effect of ``terathanisia'' that led to missing not at random in our data, with more explanation later. The dashed arrow from drug exposure to spontaneous abortion reflects a possible ``rescuing" effect that resulted in a lower spontaneous abortion rate and a higher rate of birth defects in live born fetuses. The absence of arrows from birth defects and covariates to missingness encodes our belief that spontaneous abortion contains enough information to predict missingness, which is formally imposed in Assumption \ref{assump:mar} later. 

\begin{figure}[H]
\centering
\resizebox{.9\columnwidth}{!}{
\begin{tikzpicture}
\tikzset{line width=1.5pt, outer sep=0pt,
ell/.style={draw,fill=white, inner sep=2pt,
line width=1.5pt},
swig vsplit={gap=5pt, line color right=red,
inner line width right=0.5pt}};

\node[name=D, ell, shape=ellipse] {Drug Exposure};

\node[name=Y, right=15mm of D, ell, shape=ellipse]{Birth Defects};

\node[name=M, right=15mm of Y, ell, shape=ellipse]{Spontaneous Abortion};

\node[name=O, right=15mm of M, ell, shape=ellipse]{Missingness};


\node[name=V, above=12mm of Y, ell, shape=ellipse]{Covariates};

\draw[dashed,->,line width=1.0pt,>=stealth](D) to [out=300, in=200](M);
\draw[->,line width=1.0pt,>=stealth](V) to [out=180, in=90](D);
\draw[->,line width=1.0pt,>=stealth](V) to (Y);
\draw[->,line width=1.0pt,>=stealth](V) to (M);
\draw[dashed,->,line width=1.0pt,>=stealth](D) to (Y);
\draw[->,line width=1.0pt,>=stealth](Y) to (M);

\draw[->,line width=1.0pt,>=stealth](M) to (O);

\end{tikzpicture}
}
\caption{\it A causal directed acyclic graph (DAG) describing the causal relationships between variables in the study.
}
\label{fig:dag}
\end{figure}

Table \ref{table:datasummaryoutcome} summarizes pregnancy outcomes for a total of 494 pregnant women in the OTIS data set: live birth, spontaneous abortion (SAB), therapeutic abortion (TAB), stillbirth, and lost-to-follow-up (LTFU). SAB was defined as a pregnancy loss occurring before week 20 of gestation, and stillbirth was defined as a spontaneous pregnancy loss at 20 weeks’ gestation or later. In the study sample of 494 women, 336 women were exposed to the medication during the first trimester of their pregnancy, which is the exposure window of primary interest for major birth defects, and the remaining 158 women were unexposed to the medication of interest at any time during their pregnancies. 

There were a total of 40 observed major birth defects. In addition, there were 27 missing birth defect outcomes. Twenty-five of these were  among pregnancies ending in SAB. In fact, of the 26 total SAB pregnancy outcomes, only one was observed to have a major birth defect, while the remaining 25 were missing. As mentioned earlier, SAB is known to be associated with a higher risk of major birth defects, therefore the fact that most SABs are missing this outcome falls under the mechanism of 
missing not at random (MNAR). 

\begin{table}[H]
\centering
\caption{Missing major birth defects by pregnancy outcomes}
\label{table:datasummaryoutcome}
\resizebox{0.8\columnwidth}{!}{%
\begin{tabular}{lcccccc}
  \hline
 & &{Exposed ($n=336$)} && &{Unexposed ($n=158$)} &\\ 
Birth Defect   & Yes & No & Missing & Yes & No & Missing \\ 
     \hline
  Live Birth & 30  & 287  && 5 & 139  & \\ 
  SAB & &&13 & 1 & &12  \\ 
  Stillbirth && 1  & & && \\ 
  TAB & 3    && & 1   &&\\ 
  LTFU &&&2 &&&\\
   \hline
\end{tabular}
}
\end{table}

We consider the confounders identified in the original analysis which included asthma (yes/no), maternal height and referral source. A key step in identifying the confounders was to examine whether or not the relationship between the exposure and the outcome was altered by including or excluding the covariate in question; more investigation on confounder selection for birth defect studies was carried out in \cite{xu2018impact}. The distribution of the confounders are summarized in Table \ref{table:datasummarycovariates}. More data summaries on the additional covariates are given in the Appendix.
\begin{table}[H]
\caption{Distribution of the identified confounders in OTIS study: mean (SD) or $n$ (\%).  }
\label{table:datasummarycovariates}
\centering
$ $\begin{tabular}{lcc}
  \hline
Confounders & Exposed ($n=336$) & Unexposed ($ n=158$) \\ 
  \hline
  Asthma & 45 (13.4\%) & 32 (20.3\%)\\ 
  Maternal Height (cm)  & 165 (6.98) & 167 (7.01) \\ 
   Referral Source:$^*$ &&\\
 ~~~Type I   & 26 (7.7\%)& 65 (41.1\%)\\ 
 ~~~Type II & 199 (59.2\%)& 52 (32.9\%)\\ 
  ~~~Type III & 111 (33.1\%)& 41 (26.0\%)\\ 
    \hline
\end{tabular}$ $

{\footnotesize $^*$I: TIS;  II: Pharmaceutical Company/Sponsor, Healthcare Professional; III: Patient Support Group, Internet, or Other. }
\end{table}

We apply the established mechanism of terathanasia described earlier to analyze the causal effect of the medication on major birth defects,  while handling the missing data due to SAB. We adopt the potential outcomes framework, also referred to as the Rubin causal model \citep{neyman1923applications, rubin1974, holland1986statistics}, to define the causal effects. It is important to recognize that the SAB rate was almost twice as high in the unexposed as in the exposed women. As SAB is itself a post-exposure outcome, it is possible that medication use early in gestation could have influenced the rate of pregnancy loss. It has been recognized in the literature  \citep{frangakis2002principal} that stratification based on the observed values of such a variable invalidates any causal interpretation. In this case, the previous analysis of major birth defects within the liveborn stratum did not take into account the fact that the pregnancy outcome of live birth versus SAB might be affected by exposure to the drug. For this purpose we also consider principal stratification that was proposed to address such post-exposure effects within the potential outcomes framework \citep{frangakis2002principal}.
 
The rest of the paper proceeds as follows. Due to the multiple pregnancy outcomes and complexity of the data structure, we devote the next section to describe the multiple challenges, detailed  notation, as well as the general assumptions. Section \ref{sec:ate} presents the approach we use to estimate the average treatment effect on major birth defect. Throughout the paper we will use the words `exposure' and `treatment'  inter-exchangeably. Section \ref{sec:pe} presents the principal strata as defined by the potential pregnancy outcomes, an alternative class of causal estimands that we are interested in, together with the estimation and inference approach. Section \ref{sec:realdata} provides the detailed data analysis results using the approaches described. Section \ref{sec:dis} contains conclusions and further discussion.

\section{Challenges, Notation and Assumptions}\label{sec:prep}

\subsection{Main challenges }

In the following we present the main challenges in drawing causal inference of the birth defects data, and outline our solutions to them. 

\subsubsection*{\it Missing not at random}

Birth defect presents itself before SAB but often can only be observed at the end of pregnancy. As shown in Table \ref{table:datasummaryoutcome}, the major birth defect outcome is missing for most SAB cases. Meanwhile it is well-established that in the general population, SAB cases are at higher risk of experiencing major birth defects than pregnancies that end in live births \citep{warkany1978terathansaia}. This results in missing not at random \citep[MNAR]{little2019statistical} if simply conditioning on the baseline covariates. 
Fortunately, the terathanasia theory and the temporal order of variables in Figure \ref{fig:dag} inform us how to model such a missing data mechanism. That is, to use the so-called `selection model', which models the marginal distribution of the complete data and the conditional distribution of missingness given the complete data. 
Here the complete data refer to the data with no missing major birth defect outcomes, and the missing data mechanism can be modeled using the conditional distribution of SAB given the complete major birth defect outcomes.  

\subsubsection*{\it Left Truncation and Right Censoring} 
Our SAB data is subject to left truncation, because women typically enroll in OTIS pregnancy studies after clinical recognition of their pregnancies \citep{xu:cham}. This leads to selection bias as women who have early SAB events tend not to be captured in our data. In addition, enrolled pregnancies that are subsequently lost-to-follow-up, for example, leads to right censoring. Finally, unlike death, SAB or stillbirth does not happen to all pregnancies; when a pregnancy ends in live birth, we may consider it censored for SAB/stillbirth at that time (i.e. at 20 weeks' gestation for SAB and at gestational age at delivery for stillbirth). An alternative consideration is that the pregnancy is `cured' from SAB/stillbirth if it ends in live birth; we will further discuss this later. 

Survival analysis methods have been well studied for left truncated and right-censored (LTRC)  time-to-event data. 
While likelihood based methods are often applied to left truncated survival data \citep{vardi:89, qin:etal:2011, hou2018nonparametric}, we have the dual outcome of birth defect and SAB/stillbirth. We will therefore account for the selection bias resulted from left truncation via re-weighting by the inverse of probability of not being truncated, using the conditional distribution of the enrollment time.

\subsubsection*{\it Observational nature and moderate sample size}

The prospective cohort studies in pregnancy carried out by OTIS are observational in nature. As such, the presence of confounders is inevitably an important issue to address. Many approaches exist in the literature; for pregnancy studies with birth defect outcomes in particular, various ways to select confounders and use propensity scores were discussed in \cite{xu2019statistical}. Given the numbers of events in Table \ref{table:datasummaryoutcome}, we consider parsimonious modeling approaches in the next sections, together with inverse probability weighting (IPW) using propensity scores \citep{rosenbaum1983central, d1998propensity}. This results in the minimal number of parameters that need to be estimated. 

\subsubsection*{\it Principal effects and rare events} 
As mentioned earlier, the original analysis was done on the subset of live born infants. However, live birth is a post-exposure outcome. For this data set in particular, it is possible that exposure to the drug has an effect on live birth versus otherwise, as proportionally fewer SAB's are seen in the exposed group (Table \ref{table:datasummaryoutcome}). Principal stratification is a useful framework for addressing such post-exposure complications \citep{frangakis2002principal}. The principal effects (PE) are causal effects within the principal strata, formed by the potential outcomes of the post-exposure stratification variable. This, however, does give rise to more parameters to be estimated, as compared to a single average treatment effect (ATE) over the whole population. In addition, there are parameters related to the latent membership of the principal strata. Given the already limited sample size for estimating the ATE, how to fit the PE models poses perhaps the largest challenge in this paper.

\subsection{Outcomes and notation}

Among the five types of pregnancy outcomes listed in Table \ref{table:datasummaryoutcome}, we combine SAB/stillbirth into one outcome for the purposes of this paper. As mentioned earlier, left truncation exists in our data for this variable, therefore time to SAB/Stillbirth event will be considered and survival analysis methods will be applied in order to properly handle this selection bias \citep{xu:cham, hou2018nonparametric}. In addition, TAB in pregnancy studies should be considered as a competing risk of SAB \citep{meister}, but due to the extremely low number of events in our data, it will be treated  as (non-informatively) right-censored. LTFU is the usual right censoring. Finally, live birth informs us that the pregnancy is no longer at risk of SAB/stillbirth, and it will be treated as right-censored at gestational age at delivery. For all of these survival random variables, the time scale is gestational age in weeks, and time zero is the start of gestation which is defined as the first day of the last menstrual period of a pregnant woman.

The data for subjects $i=1, ..., n$ are treated as independent and identically distributed. We first define the meanings of the following random variables:
\begin{itemize}
\item $D_i= 1$ if subject $i$ is treated or exposed, 0 otherwise;
\item $Y_i= 1$  if subject $i$ has a major birth defect, 0 otherwise, note that some $Y_i$'s are missing;
\item $O_i =1$ if $Y_i$ is observed, 0 otherwise;
\item $Q_i$  gestational age at study enrollment, i.e.~left truncation time; 
\item $T_i$ time to SAB/stillbirth; 
\item $C_i$ right censoring time, which can be TAB, LTFU or live birth delivery.
\end{itemize}
We note that live birth, for the event of interest SAB/stillbirth, is the same concept as `cured' in survival analysis. This in the literature of non-mixture cure modeling, is treated the same as censoring. It has been common practice to treat it as censoring in the analysis of SAB data; see for example \cite{meister, xu:cham, hou2018nonparametric, wu:etal}.  
In addition, TAB is competing risk for SAB \citep{meister}, but due to the small numbers here and also not to further complicate the development, we treat it as censoring here.

\begin{itemize}\item $X_i = \min(T_i, C_i)$;
\item $\Delta_i = I(T_i\leq C_i)$;
\item $M_i = 1$ if subject $i$ has an event of SAB/Stillbirth, 0 otherwise. 
Note that $1-M_i$ is the cure indicator in a mixture cure model, and $M_i$ is missing if subject $i$ is \sout{right censored} LTFU or TAB; 
\item $\boldV_i$ the vector of covariates. 
A summary of the covariates is given in the Appendix. Also let $\boldV_{nuc}$, $\boldV_{cens}$, $\boldV_{trunc}$, $\boldV_{miss}$ denote subsets of $\boldV$ that are defined and used later for conditional independence; 
\item $t_1 < t_2 < \cdots < t_K$ the $K$ distinct observed SAB/Stillbirth event times.
\end{itemize}
Note that 
other than the baseline covariates $V_i$ and exposure $D_i$, all other variables are post exposure. 

Therefore, we define the following potential outcomes of all the post exposure variables: 

\begin{itemize}
\item $(Y_i(1), Y_i(0))$:  potential major birth defect outcome under exposure or not, respectively;
\item $(M_i(1), M_i(0))$:  potential SAB/stillbirth outcome under exposure or not, respectively;
\item $(Q_i(1), Q_i(0))$:  potential gestational age at study enrollment under exposure or not, respectively;
\item $(O_i(1), O_i(0))$:  potential indicator for observed birth defect outcome under exposure or not, respectively;
\item $(T_i(1), T_i(0))$:  potential time to SAB/stillbirth under exposure or not, respectively;
\item $(C_i(1), C_i(0))$:  potential time to TAB, LTFU or live birth delivery under exposure or not, respectively.
\item $X_i(a) = \min \{ T_i(a), C_i(a) \}$, $\Delta_i(a) = I \{ T_i(a) \leq C_i(a) \}$ for $a=0, 1$.
\end{itemize} 

\subsection{Causal framework and assumptions}
We assume the following throughout the paper. The first four assumptions  are commonly adopted in causal inference.
\begin{assump}[Stable unit treatment value assumption (SUTVA)]\label{assump:SUTVA}
The potential outcomes for one subject are unaffected by the treatment assignments of other subjects, and for each subject there are no hidden versions of treatment or control being considered.
\end{assump}
The SUTVA in our case implies that whether a woman is exposed to the medication or not will not influence another woman's post-treatment outcomes like birth defects. Also the exposure window during pregnancy is well defined per study protocol.

\begin{assump}[Consistency]\label{assump:consistency}
We observe one of the potential outcomes at a time, that is, 
$Y = D\cdot Y(1) + (1 - D)Y(0)$,
$M = D\cdot M(1) + (1 - D)M(0)$,
$Q = D\cdot Q(1) + (1 - D)Q(0)$,
$O = D\cdot O(1) + (1 - D)O(0)$,
$T = D\cdot T(1) + (1 - D)T(0)$,
and
$C = D\cdot C(1) + (1 - D)C(0)$.
\end{assump}
The consistency assumption states that potential 
outcomes are equal to the factual outcomes when evaluating at the actual treatments, which allows us to link the potential outcomes to the observed outcomes. 

\begin{assump}[Positivity]\label{assump:positivity}
The propensity scores are bounded away from $0$ or $1$ given any covariates; that is, there exists $\eps > 0$ such that
\eqn
\eps \le \P(D = 1|\boldV_{nuc}) \le 1 - \eps, ~~\text{almost surely}.
\ee
\end{assump}
In the data analysis, we have $\boldV_{nuc} = \{\text{asthma, maternal height, referral sources}\}$. The assumption states that treatment assignment is not deterministic given $\boldV_{nuc}$, which allows the use of inverse probability of treatment weighting.

\begin{assump}[Conditional ignorability]\label{assump:condigno}
The treatment assignment is randomized, once given the covariates; that is,
\eqn\label{eq:ignorability}
(Y(d), O(d), T(d), C(d), M(d), V_{-nuc})\perp D~|~\boldV_{nuc},
\ee
where `$ \perp$' denotes statistical independence. 
\end{assump}
Conditional ignorability is commonly assumed in the literature which renders the  identification of causal effects possible. It requires an investigator's ability to collect sufficient covariates to ensure randomization given the covariates. The set of covariates $\boldV_{nuc}$ given above was used in the final analysis of the data set for FDA labeling purposes, and the process of confounder selection was discussed in \cite{xu2018impact}.

Assumptions \ref{assump:SUTVA} - \ref{assump:condigno} allow us to infer causal effects based on the observed data. The next two assumptions are commonly used for survival data, known as non-informative censoring and truncation. 
Note that by the theory of `terathanasia' which we adopt for  the selection model in the next section,
the birth defect outcome $Y$ is a predictor of the SAB outcome $T$. 

\begin{assump}[Conditional independent censoring]\label{assump:indcensor}
$C(d)$ is independent of $T(d)$ given $(Y(d), D=d, V_{cens})$, and there exists a finite  $\tau > 0$ such that $\P(C(d) > \tau) = 0$ and $\P(T(d) > \tau) > 0$ for $d = 0, 1$.
\end{assump}
In the analysis, 
we have  $\boldV_{cens} = \emptyset$. 

\begin{assump}[Conditional quasi-independent truncation]\label{assump:quasitrunc}
 For $d = 0, 1$, 
 $Q(d)$ and  $(T(d), C(d))$ are independent given $(D = d, \boldV_{trunc})$ on the non-truncated region, and there exists $\xi \in (0, \tau)$ such that $\P(Q(d) > \xi) = 0$.  Also $Q(d) < C(d) $ almost surely.
\end{assump}
In the analysis, 
we have $\boldV_{trunc} = \emptyset$. 

Finally, the following assumption is needed for the selection model. It states that once the information on SAB is included in the data, the major birth defect outcome is missing at random. 
\begin{assump}[Missing at random given SAB]\label{assump:mar}
\eqn\label{eq:obsatetimeindcausalmodeliptw}
O(d)  \perp (Y(d), \boldV_{-miss})~|~X(d),  \Delta(d), M(d), \boldV_{miss}.  
\ee
\end{assump}

Assumption \ref{assump:mar} is an important one in this paper, and yet not empirically verifiable. 
In the analysis, we have $\boldV_{miss} = \emptyset$. 
This is partially due to the fact that we have rare outcomes;  we note that in a selection model analysis, such a variable $\boldV_{miss}$ would be included in the model as a regressor. 
Instead, we will conduct sensitivity analysis to consider the following variables based on expert knowledge: previous SAB, previsou TAB, maternal age, maternal race, and history of a child with birth defects.


\section{Average Treatment Effect}\label{sec:ate}

In this section we describe the estimand, the average treatment effect (ATE) of the drug exposure on major birth defects and how estimation proceeds. 
\subsection{Models and Estimands}
We consider the following models for the potential outcomes:
\eqn\label{eq:yatetimeindcausalmodeliptw}
\P(Y(d) = 1) = \frac{\exp(\alpha_0 + \alpha_D \cdot d)}{1 +\exp(\alpha_0 + \alpha_D \cdot d)},
\ee
and
\eqn\label{eq:tatetimeindcausalmodeliptw}
\P(T(d) > t|Y(d) = y) = \exp\{-\Lambda(t)\exp(\beta_D \cdot d + \beta_Y \cdot y)\},
\ee
where $\Lambda(t) $ is the cumulative baseline hazard function for the conditional distribution of $T(d)$ given $Y(d)$. 
Then $\exp(\alpha_D)$ is the ATE, which is 
 the causal odds ratio 
\eqn\label{eq:averagecausaloddsratio}
\frac{P(Y(1) = 1)/P(Y(1) = 0)}{P(Y(0) = 1)/P(Y(0) = 0)}.
\ee
The parameters $\beta_D$ and $\beta_Y$ represent the effect of the treatment and birth defect on SAB/Stillbirth. In particular, $\beta_Y$ plays the role of quantifying ``terathanasia''; higher $\beta_Y$ implies a stronger effect of terathanasia.
\begin{rem}
Note that model \eqref{eq:yatetimeindcausalmodeliptw} is in fact saturated.
While we might attempt to include the interaction term between $d$ and $y$ in model \eqref{eq:tatetimeindcausalmodeliptw}, it turns out that the estimation algorithm failed to converge due to too few (7) observed birth defects ($Y = 1$) in the control group ($D = 0$), as seen from Table \ref{table:datasummaryoutcome}. 
\end{rem}

\subsection{Estimation}
We give a brief description of the estimation procedure first.
The counterfactual outcome, by definition, is not observed. In order to estimate the parameters in models \eqref{eq:yatetimeindcausalmodeliptw} and \eqref{eq:tatetimeindcausalmodeliptw}, we use the inverse probability (IP) of treatment weights to create a  pseudo-randomized sample. This will enable us to write down a weighted likelihood based on the observed variables in order to estimate the parameters in the two structural models \eqref{eq:yatetimeindcausalmodeliptw} and \eqref{eq:tatetimeindcausalmodeliptw} \citep{bres:well}. Therefore, we define the stabilized IP treatment weights as
\eqn\label{eq:ipweight}
w_i^{(1)} = \frac{\hat{\P}(D_i = 1)D_i}{\hat{\P}(D_i = 1|\boldV_{nuc, i})} + \frac{\hat{\P}(D_i = 0)(1 - D_i)}{\hat{\P}(D_i = 0|\boldV_{nuc, i})},
\ee
where $\hat{\P}(D_i = d|\boldV_i)$ is an estimated propensity score, and $\hat{\P}(D_i = d)$ is the estimated marginal probability of exposure or not.

The dataset is also subject to selection bias introduced by truncation. 
We use inverse probability of truncation weights to account for it. 
Define the IP truncation weights as
\eqn\label{eq:trweight}
w_i^{(2)} = \frac{1}{\hat{\P}(Q_i < q|D_i, \boldV_{trunc, i})|_{q = T_i}}.
\ee

The estimation of the weights will be described later in the data analysis.

Write $\balpha = (\alpha_0, \alpha_D)^\top$, $\bbeta = (\beta_D, \beta_Y)^\top$, and 
$\btheta = (\balpha^\top, \bbeta^\top, \lambda_1, ..., \lambda_K)^\top$ from \eqref{eq:yatetimeindcausalmodeliptw} and \eqref{eq:tatetimeindcausalmodeliptw}, 
where $ \lambda_1, ..., \lambda_K$ are the point masses at $t_1, ..., t_K $ after discretizing the baseline hazard function according to the semiparametric theory. 
The weighted likelihood based on the complete data, including the augmented $Y_i$'s if they are missing,  can then be written: 
\eqn\label{eq:ateweightcomplike}
L_w^c(\btheta) 
= \prod_{i = 1}^n \Big\{\pi_i^{Y_i}(1 - \pi_i)^{1 - Y_i}\lambda_i(X_i)^{\Delta_i}  S_i(X_i)\Big\}^{w_i},
\ee
where $w_i = w_i^{(1)}w_i^{(2)} $,
$\pi_i = \expit(\alpha_0 + \alpha_DD_i)$,
$\lambda_i(t) = \lambda_0(t)\exp(\beta_DD_i + \beta_YY_i)$ with $\lambda_0(\cdot)$ equal to the corresponding $\lambda_k$, and $S_i(t) = \exp\{-\Lambda(t)\exp(\beta_DD_i + \beta_YY_i)\}$ with $ \Lambda(\cdot) $ equal to the corresponding cumulative sum of the $\lambda_k$'s. This yields a weighted complete data log-likelihood:
\eqn\label{eq:ateweightcomploglikeoriginal}
&& l_w^c(\btheta) = \log L_w^c(\btheta) \nonumber\\
&=&  \sum_{i = 1}^n w_i^{TR}w_i^{IP}\Big[Y_i\log \pi_i + (1 - Y_i)\log(1 - \pi_i) + \Delta_i \log \lambda_i(X_i) + \log S_i(X_i) \Big\} \Big].
\ee
We use the following ES algorithm to maximize the weighted complete data log-likelihood. 

\subsubsection{Weight Computation}
We first estimate the conditional survival function of the truncation time $Q$. This can be done by imposing a model on the conditional distribution function of $Q$, evaluated at $T_i$ given $V_i$ and $D_i$. This is compatible with our marginal structural model by Assumption \ref{assump:quasitrunc}. We then compute the weights $w_i^{(1)}$ for $1 \le i \le n$.  $\hat{\P}(D_i = d|\boldV_i)$ in the denominators of \eqref{eq:ipweight} can be obtained using any propensity score approaches,
for example the R package ``twang'' \citep{twang}, for $d = 0, 1$.

\subsubsection{ES algorithm}

To estimate the parameters $\theta$ in the weighted complete data log-likelihood  
\eqref{eq:ateweightcomploglikeoriginal}, we use an expectation-substitution (ES) algorithm, which is very similar to the well-known EM algorithm but for estimating equations instead of likelihood functions \citep{rosen:etal, elashoff2004algorithm}. Note that the weighted likelihood approach is in fact an Z-estimation approach, leading to sandwich type variance estimates instead of inverse of the Fisher information. 
Both the EM and the ES algorithms have since been applied to nonparametric (weighted) likelihoods under semiparametric models  \cite[for example]{peng2000nonparametric, vaid:xu:00, faig2013}. 
In the following we described the ES algorithm as applied to \eqref{eq:ateweightcomploglikeoriginal}.

\medskip
\noindent {\it Initialization}
\medskip

We initiate $\theta^{(0)}$ by treating all the missing  $Y$ values as $0$, ignoring left truncation, and fitting a logistic and a Cox regression model corresponding to \eqref{eq:yatetimeindcausalmodeliptw} and \eqref{eq:tatetimeindcausalmodeliptw}, respectively, with weights $w_i$. We fit using `glm()' and `coxph()' functions in R with weights.

\medskip
\noindent {\it E-step}
\medskip

At step $t+1$, for $t=0, 1, ...$,  let $Q(\btheta|\btheta^{(t)}) = \E_{\btheta^{(t)}}[l_w^c(\btheta)|\cO]$, where $\cO$ denotes all the observed variables. We have
\eqn
&&Q(\btheta|\btheta^{(t)}) = \sum_{i = 1}^{n}\Big[w_{1, i}^\pi\log \pi_i + w_{0, i}^\pi \log (1 - \pi_i) \nonumber\\ 
&&+ \sum_{k = 1}^K w_{i, k, 1}^f\log f_i(t_k|Y_i = 1) + \sum_{k = 1}^K w_{i, k, 0}^f\log f_i(t_k|Y_i = 0)\nonumber\\
&& + w_{i,1}^S \log S_i(X_i|Y_i = 1) + w_{i, 0}^S \log S_i(X_i|Y_i = 0)\Big] \label{eq:ateqfunc},
\ee
where
\eqn\label{w_1i}
w_{1, i}^\pi &=& w_i \{O_iY_i + (1 - O_i)\P_i^{(t)}(Y_i^{mis} = 1)\},\\
w_{0, i}^\pi &=& w_i \{O_i(1 - Y_i) + (1 - O_i)\P_i^{(t)}(Y_i^{mis} = 0)\},\\
w_{i, k, 1}^f &=& w_i (O_iY_i +  (1 - O_i)\P_i^{(t)}(Y_i^{mis} = 1))\nonumber\\
&&\cdot[\Delta_i\mathbbm{1}(X_i = t_k)],\\
w_{i, k, 0}^f &=& w_i (O_i(1 - Y_i) +  (1 - O_i)\P_i^{(t)}(Y_i^{mis} = 0))\nonumber\\
&&\cdot[\Delta_i\mathbbm{1}(X_i = t_k)],\\
w_{i, 1}^S &=& w_i (1 - \Delta_i)[ O_iY_i + (1 - O_i)\P_i^{(t)}(Y_i^{mis} = 1)], \\
\label{w_i0}
w_{i, 0}^S &=& w_i (1 - \Delta_i)[ O_i(1 - Y_i) + (1 - O_i)\P_i^{(t)}(Y_i^{mis} = 0)].
\ee
The expressions for $\P_i^{(t)}(Y_i^{mis} = y)$, $\E_i^{(t)}(A_i|Y_i = y)$ and $\P_i^{(t)}(T_{i1} = t_k|Y_i = y)$ in the above are given in the Appendix.

\medskip
\noindent {\it S-step}
\medskip

In the S-step we update $\theta$ as the maximizer of $Q(\btheta|\btheta^{(t)})$. It is seen that $Q(\btheta|\btheta^{(t)})$ can be decomposed as:
\eqn
Q(\btheta|\btheta^{(t)}) = l_{glm}(\balpha) + l_{cox}(\bbeta, \lambda_1, \cdots, \lambda_K),
\ee
where
\eqn\label{l_glm}
l_{glm}(\balpha) = \sum_{i = 1}^n\Big[w_{1, i}^\pi\log \pi_i + w_{0, i}^\pi \log (1 - \pi_i)\Big],
\ee
and
\eqn\label{l_cox}
l_{cox}(\bbeta, \lambda_1, \cdots, \lambda_K) &=& \sum_{i = 1}^n\Big\{\sum_{k = 1}^K\Big[w_{i, k, 1}^f\log f_i(t_k|Y_i = 1) + \sum_{k = 1}^K w_{i, k, 0}^f\log f_i(t_k|Y_i = 0)\Big]\nonumber\\
&& + w_{i, 1}^S \log S_i(X_i|Y_i = 1) + w_{i, 0}^S \log S_i(X_i|Y_i = 0)\Big\}.
\ee
Consequently, we fit two weighted regression models, using the weights in \eqref{l_glm} and \eqref{l_cox} that are computed in \eqref{w_1i} - \eqref{w_i0} of the E-step above. This can again be done using `glm()' and `coxph()' functions in R.  

The E-step and the S-step are iterated until convergence, which may be declared when the change in the consecutive parameter values is below a pre-specified threshold; in this paper we use $L^2$ norm $<0.00001$. 
The asymptotic properties of the estimator was studied in \cite{ying2020statistical}, which showed that the estimator is consistent and asymptotically Gaussian.
We estimate the variance of our estimator via the multiplier bootstrap \citep{kosorok}. The multiplier bootstrap, in each time of repetition, weigh each individual by an independent standard exponential random variable. We then repeat our estimation for $B = 200$ times with weights to get bootstrapped estimates. The variance of 200 bootstrapped estimates is used to estimate the variance of our estimator. We use multiplier bootstrap because of its stability in moderate sample sizes compared to the nonparametric resampling bootstrap.


\subsection{Simulation}

Here we carry out a small simulation experiment to examine the validity of our estimation and inferential procedure. We simulate a randomized dataset with left truncation, right censoring and missing outcomes. To keep the data generating process simple we do not include any baseline covariates.
The simulation details are given in the Appendix, and the results are presented in Table \ref{tab:simu}, where `SE' is the average of estimated standard errors, `SD' is the standard deviation of the estimates from the 500 simulation runs, and `CP' is coverage probability of the nominal 95\% confidence intervals.

\begin{table}[H]
\caption{Simulation results}
\label{tab:simu}
$$\begin{tabular}{lcccc}
  \hline
 &Bias& SE & SD& CP (\%)\\ \hline
$\alpha_0$ &-0.021 &0.162&0.166&95.0\\
$\alpha_D$ &0.007 &0.222&0.234&95.2\\
$\beta_D$ &0.002 &0.104&0.107&96.0\\
$\beta_Y$ &-0.010 &0.117&0.120&95.6 \\
\hline
\end{tabular}$$
\end{table}

\section{Principal Effects}\label{sec:pe}

In this section we describe an alternative class of estimands called ``principal effects''. This serves as a separate analysis from the previous ATE. As explained earlier, the current practice of stratifying on live born or not does not have causal interpretation, because it is a post exposure variable. Instead, principal effects as described below carry causal meanings. 

\subsection{Models and Estimands}
As mentioned earlier SAB/stillbirth is a post exposure variable, therefore as we defined before  it can take on two potential values $M_i(0)$ and $M_i(1)$ for subject $i$. As explained in \cite{frangakis2002principal} a stratified comparison of the $Y_i$'s based on the observed values of the $M_i$'s, is equivalent to comparing $ \P( Y_i(1)=1 | M_i(1)=m )$ versus $ \P( Y_i(0)=1 | M_i(0)=m )$. Such a comparison is problematic because the set of subjects $\{i: M_i(1)=m\} $ is not the same set of subjects $\{i: M_i(0)=m\} $, as long as the exposure has non-zero effect on SAB/stillbirth. Unfortunately this is likely the case here, leading to so-called post-treatment selection bias in the estimated exposure effect \citep{rosenbaum1984consequences, robins1992identifiability, frangakis2002principal}.

In this section we consider principal stratification which is the stratification with respect to the joint potential values of $M$. Namely, we use $(M(0), M(1))$ to stratify the whole population. The whole population is then divided into:
\begin{enumerate}
    \item Always-survivors (SS), $(M(0), M(1)) = (0, 0)$, are those subjects who will not experience SAB/stillbirth no matter whether  treated or not;
    \item Treatment-survivors (NS), $(M(0), M(1)) = (1, 0)$, are those subjects who will experience SAB/stillbirth only when not  treated;
    \item Control-survivors (SN), $(M(0), M(1)) = (0, 1)$, are those subjects who will experience SAB/stillbirth only when  treated;
    \item Never-survivors (NN), $(M(0), M(1)) = (1, 1)$, are those subjects who will experience SAB/stillbirth no matter  treated or not.
\end{enumerate}
 Table \ref{table:psdef} shows the division of the whole population into the above four principal strata.
\begin{table}[H]
\centering
\caption{Division of the whole population into four principal strata }
\label{table:psdef}
$ $\begin{tabular}{|c|c|c|c|}
\hline
\multicolumn{2}{|c|}{\multirow{2}{*}{}} & \multicolumn{2}{c|}{M(1)}  \\ \cline{3-4} 
\multicolumn{2}{|c|}{}                  & 0           & 1            \\ \hline
\multirow{2}{*}{M(0)}        & 0        & Always-survivors (SS)  & \cellcolor{gray!25} Control-survivors (SN)  \\ 
\cline{2-4} 
                             & 1        &  Treatment-survivors (NS)   & Never-survivors (NN) \\ 
                             \hline
\end{tabular}$ $
\end{table}

Due to the very limited number of events in our data, in the following we further make a  monotonicity assumption that eliminates the `control-survivor' stratum, in order to reduce the number of parameters that need to be estimated later. 
We assume
\begin{assump}[Monotonicity]\label{assump:monotonicity}
$M(1) \le M(0)$ with probability one. 
\end{assump}
For our data, this means that a woman is less likely to have an SAB event under exposure to the medication than otherwise.
This is the supported by the empirical data as well as the fact that the estimated $\beta_D$ under model \eqref{eq:tatetimeindcausalmodeliptw} is negative (see Section \ref{sec:realdata}); that is, 
the drug reduces the risk of SAB/stillbirth. 

It is unknown which principal stratum a subject belongs to. 
However, certain relationship can be derived between the latent principal strata and
the observed group of subjects defined according to $(D, M^{obs})$, where $M^{obs} = M$ if observed, and $M^{obs} =~?$ otherwise. Table \ref{table:observedgroup} summarizes the correspondence between the observed groups and the latent strata. For example, those with $ (D_i, M_i^{obs}) = (0, 0)$ i.e.~no SAB/stillbirth events under no treatment, can only belong to the always-survivor stratum SS due to the monotonicity assumption. On the other hand, those with $ (D_i, M_i^{obs}) = (0, 1)$ i.e.~having had SAB/stillbirth events under no treatment, can belong to either NS (treatment-survivors) or NN (never-survivors). Missing $M_i$ leads to possibilities of all three strata, etc. Table \ref{table:observedgroup} also gives the number of subjects (group size) and the number of birth defects in each observed group for the OTIS  data. 
\begin{table}[H]
\centering
\caption{Correspondence between the observed $O(D, M^{obs})$ groups and the latent principal strata}
\label{table:observedgroup}
$ $\begin{tabular}{ccccl}
\hline
$O(D, M^{obs})$ & Size  & Birth Defects &Missing Defects & Principal Strata \\ \hline
$O(0, 0)$       &   144              &5  &0  & SS \\ 
$O(0, 1)$       &13                  &1 &12    & NS, NN  \\
$O(0, ?)$       &1                 &1  &0  & SS, NS, NN  \\
$O(1, 0)$       &    317              &  30 & 0   & SS, NS         \\
$O(1, 1)$        & 14    &0      &13           & NN  \\ 
$O(1, ?)$        & 5    &3       &  2        & SS, NS, NN  \\  
\hline
\end{tabular}$ $
\end{table}

Following \cite{frumento2012evaluating}
we define $G$ as the latent indicator for the principal strata, which takes values in $\{\SS, \NS, \NN\}$. We assume a multinomial distribution for the principal strata membership:
\eqn\label{eq:psstrata}
\P(G = g) = \frac{\exp(\gamma_g)}{\sum_{g'}\exp(\gamma_{g'})},
\ee
where $g \in \{ \SS, \NS, \NN \}$, and we treat the group SS as reference,  i.e.~$\gamma_{\SS} = 0$. 
Parallel to model \eqref{eq:yatetimeindcausalmodeliptw} for the ATE in Section \ref{sec:ate}, the causal estimands are now the principal effects $\alpha_{D, g}$ in each stratum: 
\eqn\label{eq:psycausalmodel}
\P(Y(d) = 1|G = g) = \frac{\exp(\alpha_{0, g} + \alpha_{D, g} \cdot d)}{1 + \exp(\alpha_{0, g} + \alpha_{D, g} \cdot d)}.
\ee

The parameters of scientific interest in models like the above can vary depending on the applications. In a somewhat similar-in-appearance but different setting referred to as `truncation by death', it is often argued that the only stratum to be considered is equivalent to our `always-survivors', and the principal effect there is  referred to as the `survivor average causal effect' (SACE) \citep{ding2011identifiability, yang2016using, ding2018causal}. We further discuss the difference in our setup in the last section of the paper. In the context of drug exposure during pregnancy, we can be potentially interested in the principal effects in all strata. This is at least partially due to the scientific need to understand the drug mechanism in causing birth defects (or not) in both live born infants as well as in fetuses that are lost due to SAB/stillbirth. Furthermore, it can be helpful for some women with a pregnancy loss to know that the fetus was malformed. It will be seen, however, due to the limited number of events in our data, we may not be able to reliably estimate all the parameters in \eqref{eq:psycausalmodel}. Eventually we can only draw valid inference of a subset of parameters.

Finally in order to handle left truncation in the data, as before we consider the potential time to SAB/Stillbirth $T(d)$, but now in each of the three   principal strata. We note that $T(d)= \infty$ in SS, and also in NS if $d = 1$. On the other hand, $T(d) < \infty$ in NN, and in NS if $d=0$. For these latter three cases where $T(d) < \infty$ we assume:
\eqn
&&\P(T(d) > t|Y(d) = y, G = g) \nonumber\\
&=& \exp\left[ - \Lambda(t)\exp\{ \beta_{0, \NS}\cdot (1 - d)\cdot \mathbbm{1}(g = \NS) + \beta_{D, \NN}\cdot d \cdot \mathbbm{1}(g = \NN) + \beta_{Y} \cdot y \} \right]\label{eq:pstcausalmodel}.
\ee
Note that the intercept $\beta_{0, \NN}$ is absorbed  in the baseline cumulative hazards $\Lambda(t)$, and thus we only have $\beta_{0, \NS}$ and $\beta_{D, \NN}$ in the model above. 

We note that the survival  model \eqref{eq:pstcausalmodel} only concerns the timing of the events among those who are susceptible. This is very similar to the `cure' model concept \citep{farewell, farewell:86, kuk1992mixture, sy:taylor, lu:ying, hou2018nonparametric}.
The effect of drug exposure  on the occurrence of SAB/stillbirth, on the other hand, is now reflected in the sizes of the principal strata. In particular, we may estimate the causal log odds ratio  \citep{frumento2012evaluating}
\eqn\label{eq:treatmentsabps}
\log(\mbox{OR}_M) = \log\bigg\{\frac{\P(M(1) = 1)/ \P(M(1) = 0)}{ \P(M(0) = 1)/ \P(M(0) = 0)}\bigg\}
\ee
by taking logarithm of
\eqn
\widehat{\mbox{OR}_M } &=&\frac{\hat \P(M(1) = 1)/\hat \P(M(1) = 0)}{\hat \P(M(0) = 1)/\hat \P(M(0) = 0)} \nonumber \\
&=&\frac{\hat \P(G = \NN)\big/[\hat \P(G = \NS) + \hat \P(G = \SS)]}{[\hat \P(G = \NN) + \hat \P(G = \NS)]\big/ \hat \P(G = \SS)}\nonumber\\
&=& \frac{\exp(\hat \gamma_{\NN})}{ \{\exp(\hat \gamma_{\NS}) + \exp(\hat \gamma_{\NN})\}\{1 + \exp(\hat \gamma_{\NS})\} }. \label{eq:treatmenteffectsabexpress}
\ee
This quantity encodes the causal effect of the prenatal drug on spontaneous abortion. A possible value reflects that the drug increases the chance of a woman to experience SAB. Note that the second line above made use of the monotonicity assumption that $\P(G = \SN) = 0$.

\subsection{Estimation}

The parameterization with mixture of populations and more parameters for the principal effects leads to a highly non-convex as well as flatter surface of the weighted observed data log-likelihood. In this case, an algorithm like the ES in Section \ref{sec:ate}, which indirectly works on the weighted observed data log-likelihood and only guarantees to find a local optima, is insufficient for optimization. Directly applying the ES algorithm in this case either causes the parameters to diverge or converge to a local optima. To overcome this challenge, 
a more granular optimization algorithm that is designed to search for the global optima is needed. To that end, we adopt ``Improved Stochastic Ranking Evolution Strategy'' \citep{runarsson2005search} implemented in the ``nloptr'' package \citep{nloptr} available on R CRAN, which can directly trace the value of the objective function (i.e.~the weighted observed data log-likelihood). 
We can trace the value of the objective function while monitoring the parameters of interest. We stop the optimization when the parameters of interest converge.

The weighted observed data likelihood is
\eqn\label{eq:pelik}
L_w^o(\btheta) 
&=& \prod_{i \in O(0, 0)} \Big[p_{\SS}\pi_i^{Y_i}(1 - \pi_i)^{1 - Y_i} \Big]^{w_i} \nonumber \\
&&\prod_{i \in O(0, 1)} \Big[p_{\NS}\pi_i^{Y_i}(1 - \pi_i)^{1 - Y_i}\lambda_i(X_i) S_i(X_i) + p_{\NN}\pi_i^{Y_i}(1 - \pi_i)^{1 - Y_i}\lambda_i(X_i) S_i(X_i)\Big]^{w_i \mathbbm{1}(O_i = 1)} \nonumber \\
&&\prod_{i \in O(0, 1)} \Big[p_{\NS}(\pi_i^{Y_i}\lambda_i(X_i) S_i(X_i) + (1 - \pi_i)^{1 - Y_i}\lambda_i(X_i) S_i(X_i)) \nonumber \\
&&~~+ p_{\NN}(\pi_i^{Y_i}\lambda_i(X_i) S_i(X_i) + (1 - \pi_i)^{1 - Y_i}\lambda_i(X_i) S_i(X_i))\Big]^{w_i \mathbbm{1}(O_i = 0)} \nonumber \\
&&\prod_{i \in O(0, ?)} \Big[p_{\SS}\pi_i^{Y_i}(1 - \pi_i)^{1 - Y_i} + p_{\NS}\pi_i^{Y_i}(1 - \pi_i)^{1 - Y_i}S_i(X_i) + p_{\NN}\pi_i^{Y_i}(1 - \pi_i)^{1 - Y_i}S_i(X_i) \Big]^{w_i } \nonumber \\
&&\prod_{i \in O(1, 0)} \Big[p_{\SS}\pi_i^{Y_i}(1 - \pi_i)^{1 - Y_i} + p_{\NS}\pi_i^{Y_i}(1 - \pi_i)^{1 - Y_i}\Big]^{w_i } \nonumber \\
&&\prod_{i \in O(1, 1)} \Big[p_{\NN}\pi_i^{Y_i}(1 - \pi_i)^{1 - Y_i}\lambda_i(X_i)  S_i(X_i) \Big]^{w_i \mathbbm{1}(O_i = 1)} \nonumber \\
&&\prod_{i \in O(1, 1)} \Big[p_{\NN}(\pi_i^{Y_i}\lambda_i(X_i) S_i(X_i) + (1 - \pi_i)^{1 - Y_i}\lambda_i(X_i) S_i(X_i)) \Big]^{w_i \mathbbm{1}(O_i = 0)}\\
&&\prod_{i \in O(1, ?)} \Big[p_{\SS}\pi_i^{Y_i}(1 - \pi_i)^{1 - Y_i} + p_{\NS}\pi_i^{Y_i}(1 - \pi_i)^{1 - Y_i} + p_{\NN}\pi_i^{Y_i}(1 - \pi_i)^{1 - Y_i} S_i(X_i) \Big]^{w_i \mathbbm{1}(O_i = 1)} \nonumber \\
&&\prod_{i \in O(1, ?)} \Big[p_{\SS} + p_{\NS} + p_{\NN}(\pi_i^{Y_i}S_i(X_i) + (1 - \pi_i)^{1 - Y_i}S_i(X_i)) \Big]^{w_i \mathbbm{1}(O_i = 0)}.
\ee

\section{OTIS Data Analysis Results}\label{sec:realdata}

We use the R package ``twang'' \citep{twang} to estimate the propensity scores  $\hat \P(D_i = 1|\boldV_{nuc, i})$ with the previously described confounders, which are then used to form the stabilized  weights $w_i^{(1)}$ as in \eqref{eq:ipweight}. 
In the Appendix we show that all the confounders are well balanced after weighting. 

\begin{figure}
\centering
\includegraphics[scale=0.4]{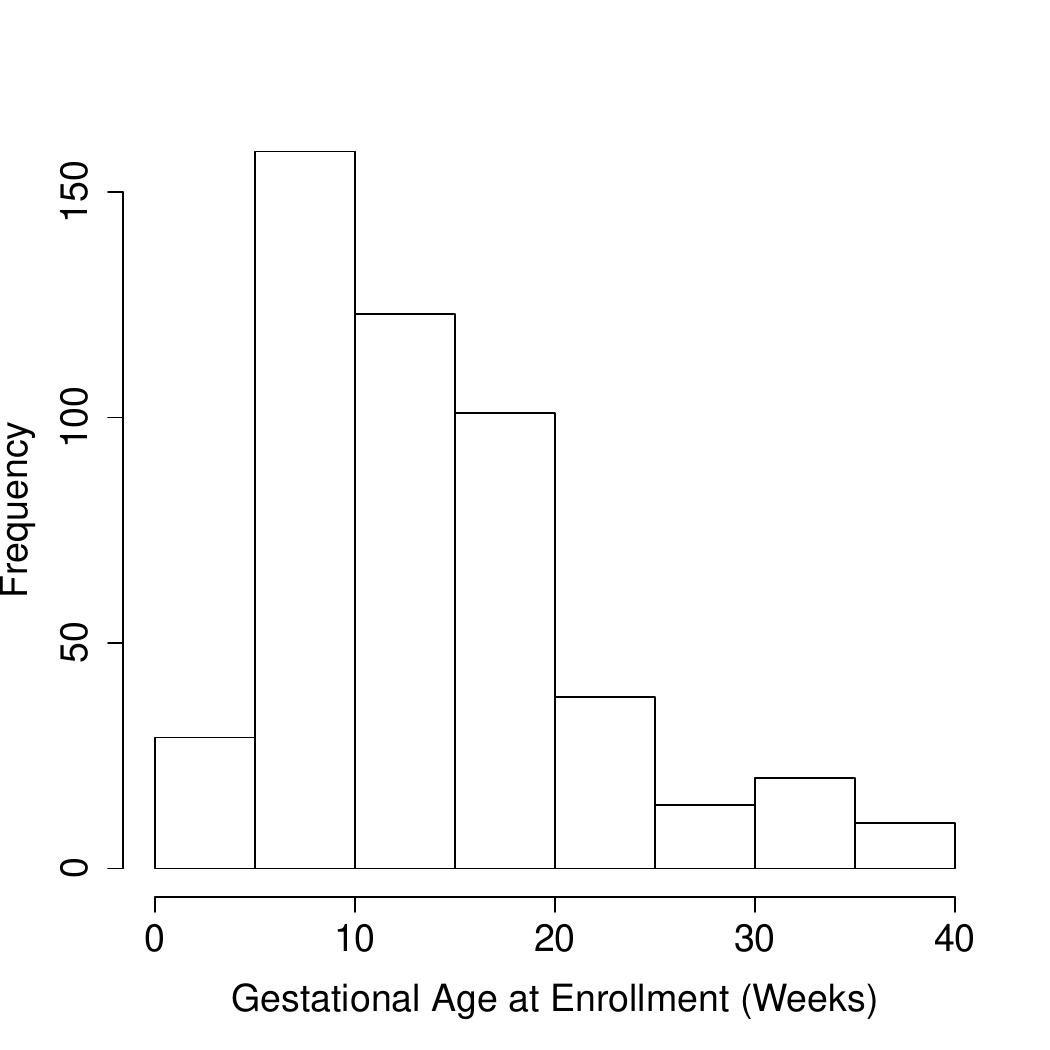}
\includegraphics[scale=0.4]{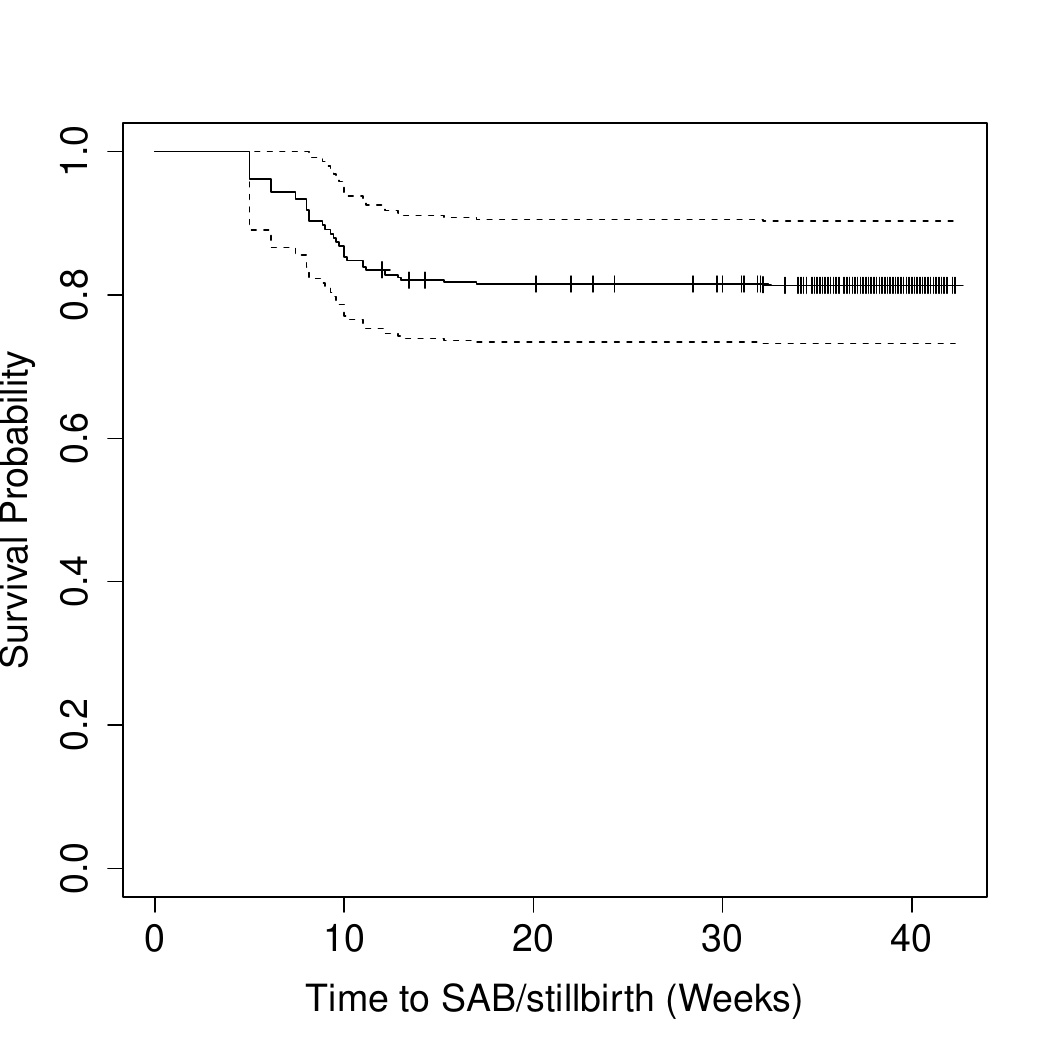}
\includegraphics[scale=0.4]{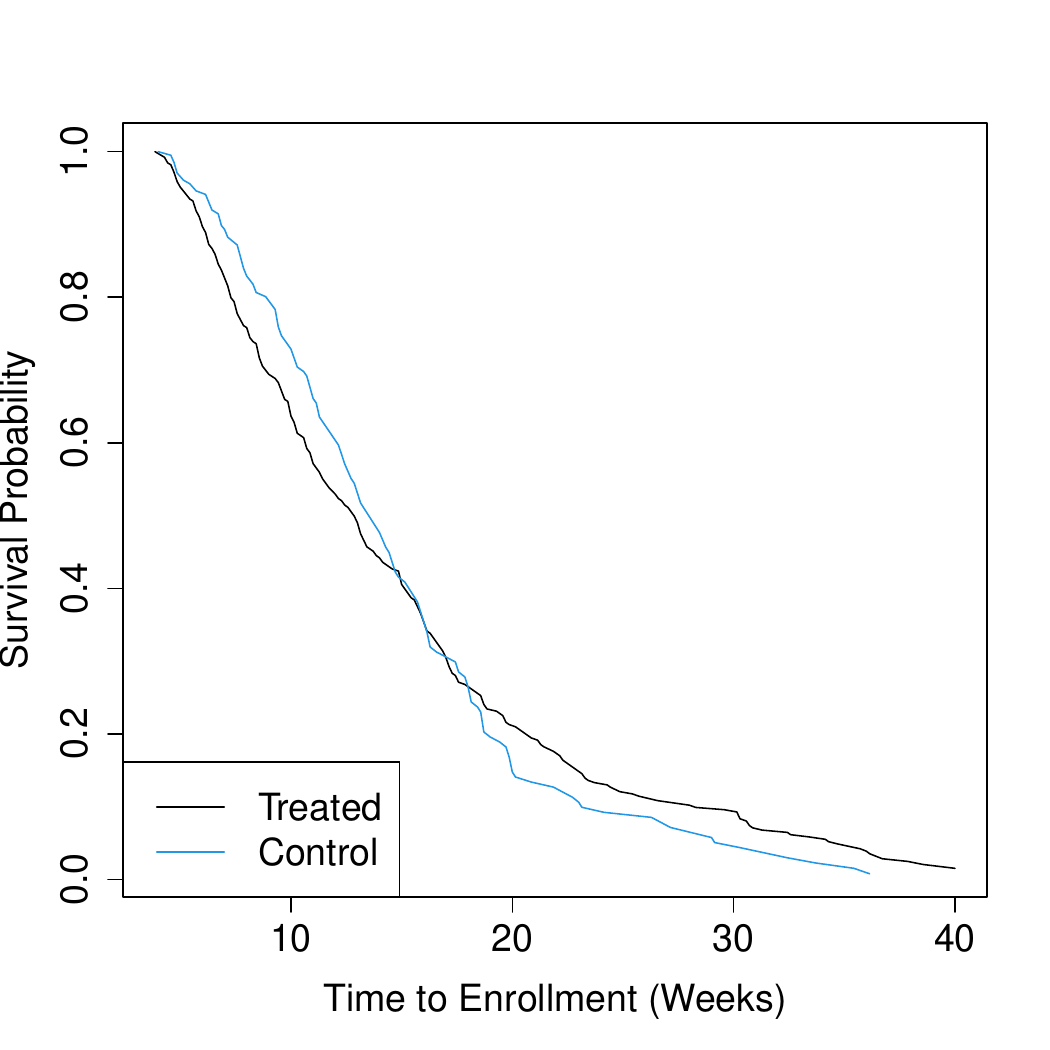}
\includegraphics[scale=0.4]{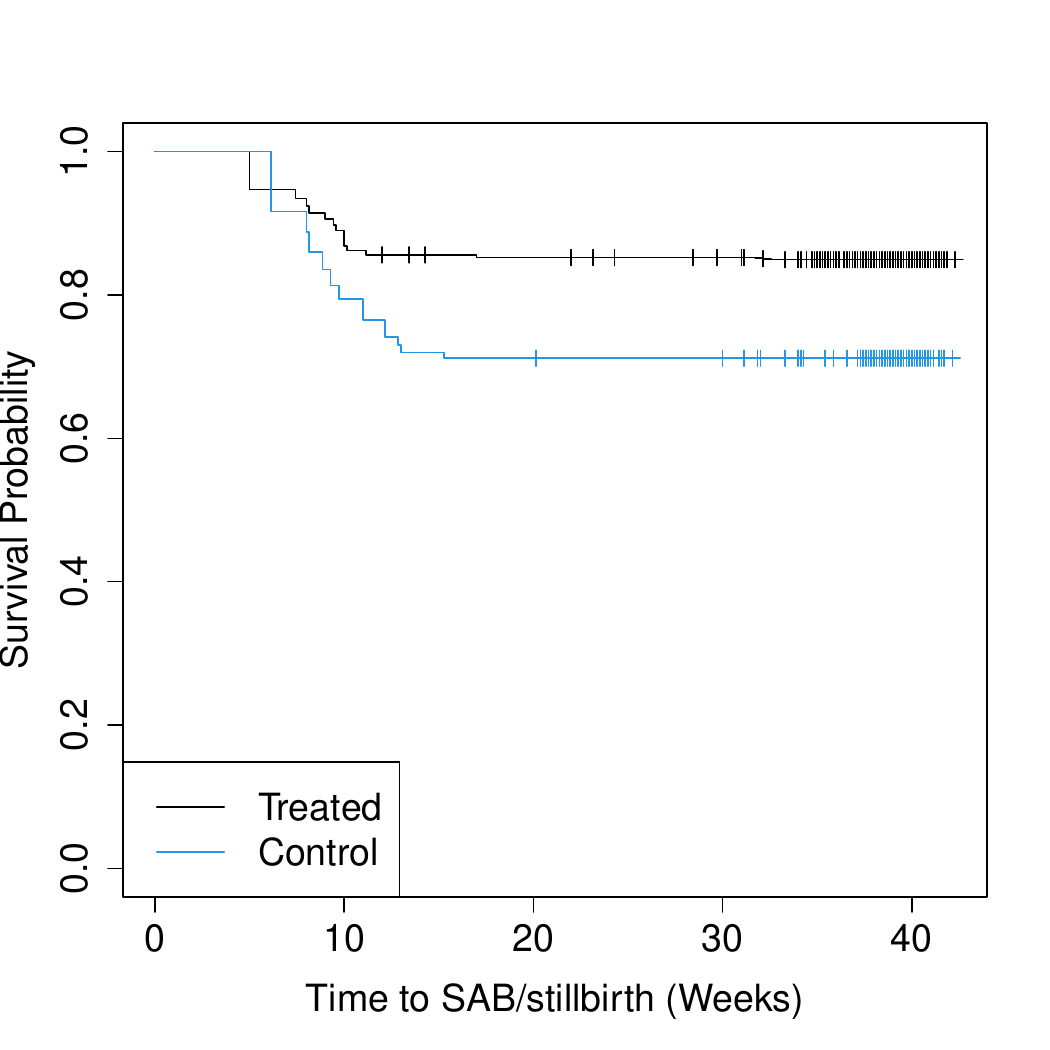}
\caption{\it Histogram of gestational age at study enrollment (top left), KM estimate for time to SAB/stillbirth with confidence intervals (top right), product-limit estimates for enrollment time by treatment groups (bottom left), and KM estimates for time to SAB/stillbirth by treatment groups (bottom right).}
\label{fig:sabtreatment}
\end{figure}

Figure \ref{fig:sabtreatment} shows the histogram of the left truncation time $Q$, the Kaplan-Meier (KM) estimate of the time to SAB/stillbirth $T$ distribution accounting for left truncation by $Q$ and right censoring by $C$, as well as the product limit estimate of the survival curves for $Q$ accounting for right truncation by $X$. Note that right truncation of $Q$ by $T$ is equivalent to right truncation by $X$ under Assumption \ref{assump:quasitrunc}, and right truncation can be handled by left truncation techniques using a transformed time scale such as  $-Q$ and $-X$. The estimated $Q$ distribution by treatment group is then used to form the inverse truncation weights $w_i^{(2)}$  as in \eqref{eq:trweight}.

\subsection{Results on Average Treatment Effect}\label{sec:ateresult}

The estimates of the parameters in \eqref{eq:yatetimeindcausalmodeliptw} and \eqref{eq:tatetimeindcausalmodeliptw} and their  standard errors (SE) etc.~are presented in Table \ref{table:atetimeindresult}. From the table we see that the estimated $\hat \beta_Y >0$ is consistent with the known ``terathanasia'' theory mentioned before. However, it has a very wide 95\% confidence interval (CI), which is perhaps not surprising as we only had two observed major birth defect outcome $Y_i$'s among the 27 SAB/stillbirth events. The results otherwise show that the drug has a nonsignificant effect in increasing major birth defects, with a causal odds ratio less than 2. In addition, exposure to the drug has a negative albeit not significant effect on SAB/Stillbirth, reducing the hazard to less than half of the unexposed.
\begin{table}
\centering
\caption{Parameter estimates from the OTIS data for ATE analysis}
\label{table:atetimeindresult}
$ $\begin{tabular}{lrccc}
\hline
& Estimate (SE) & exp(Estimate) & 95\% CI of OR/HR  & $p$-value \\ \hline
$\alpha_0$ & -2.184  (0.871) & 0.113 & (0.020, 0.629) & 0.012 \\ 
$\alpha_D$ & 0.563  (0.718) &1.755 & (0.430, 7.172) & 0.433 \\ 
$\beta_D$ & -0.750  (0.424) & 0.472 & (0.206, 1.083) & 0.076 \\ 
$ \beta_Y$ & 2.245  (2.123) & 9.440 & (0.147, 605.5) & 0.290 \\   \hline
\end{tabular}$ $
\end{table}

\begin{figure}
\centering
\includegraphics[scale=0.6]{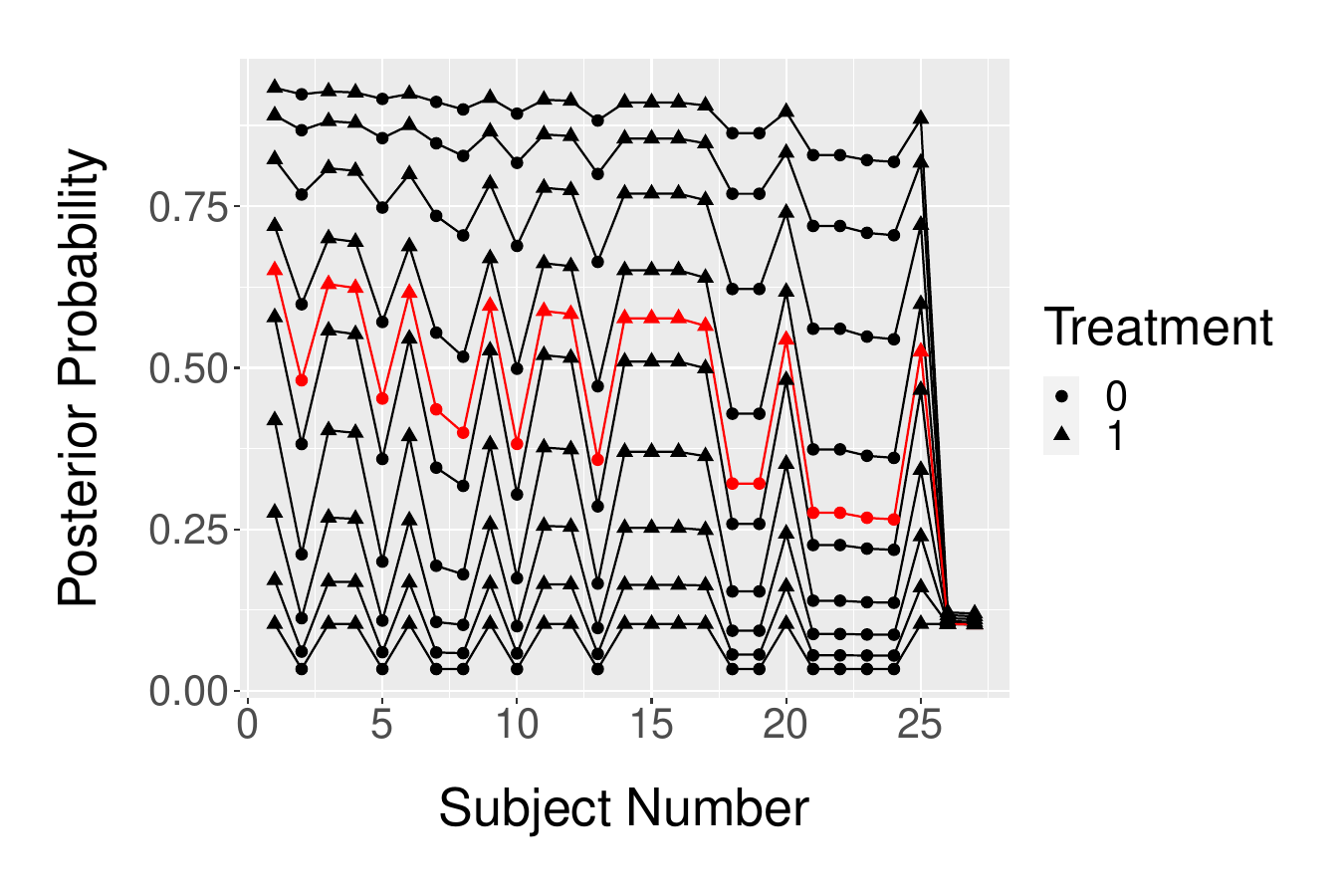}
\caption{\it Posterior probabilities of major birth defect as a function of $\beta_Y$ for subjects with missing major birth defect outcomes, ordered by possibly censored event time. The red line is when $\beta_Y$ is set at the estimated $\beta_Y$ value, black lines from bottom to top correspond to $\beta_Y$ set from 0 to 4 with increment of 0.5. The last two subjects are censored.
}
\label{fig:atetimeindpY}
\end{figure}

Due to the little confidence we have in the estimated $\hat \beta_Y$ as reflected in its wide CI, we further conduct a sensitivity analysis to examine the robustness of our conclusion about the ATE $\alpha_D$ with respect to the value of $\beta_Y$, which affects the probability of major birth defects among those with missing values. Following the theory of terathanasia, we restrict $\beta_Y$ to be non-negative, and  vary it on the interval $[0, 4]$; note that 4 would be considered an extremely large log hazard ratio. Figure \ref{fig:atetimeindpY} shows the posterior probabilities of major birth defect among  the 27 subjects (ordered  by their SAB event times) with missing values, which increase as $\beta_Y$ becomes larger.  We also note that the probabilities of birth defect are generally higher for the subjects from the exposed group (triangles) than the unexposed group (circles). 

\begin{figure}[H]
\centering
\includegraphics[scale=0.425]{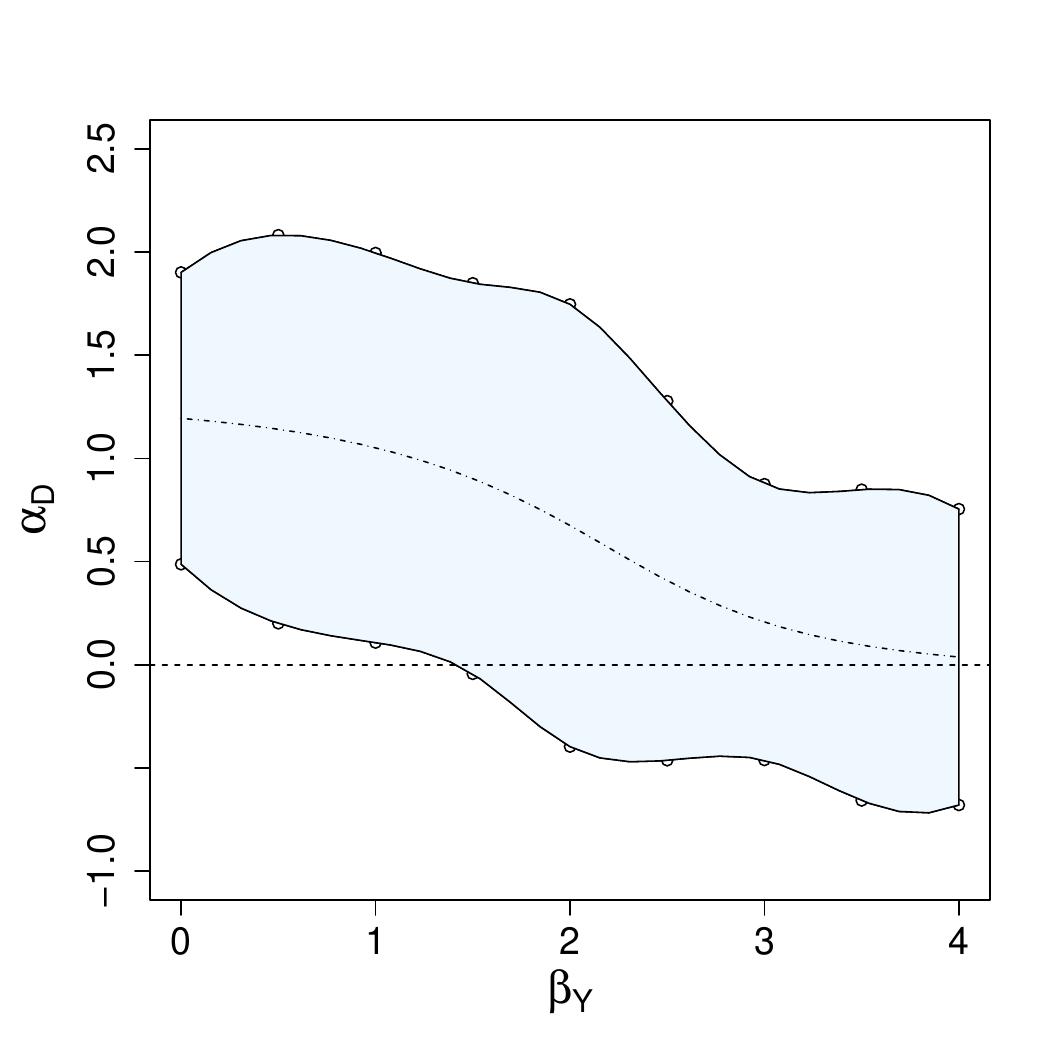}
\includegraphics[scale=0.425]{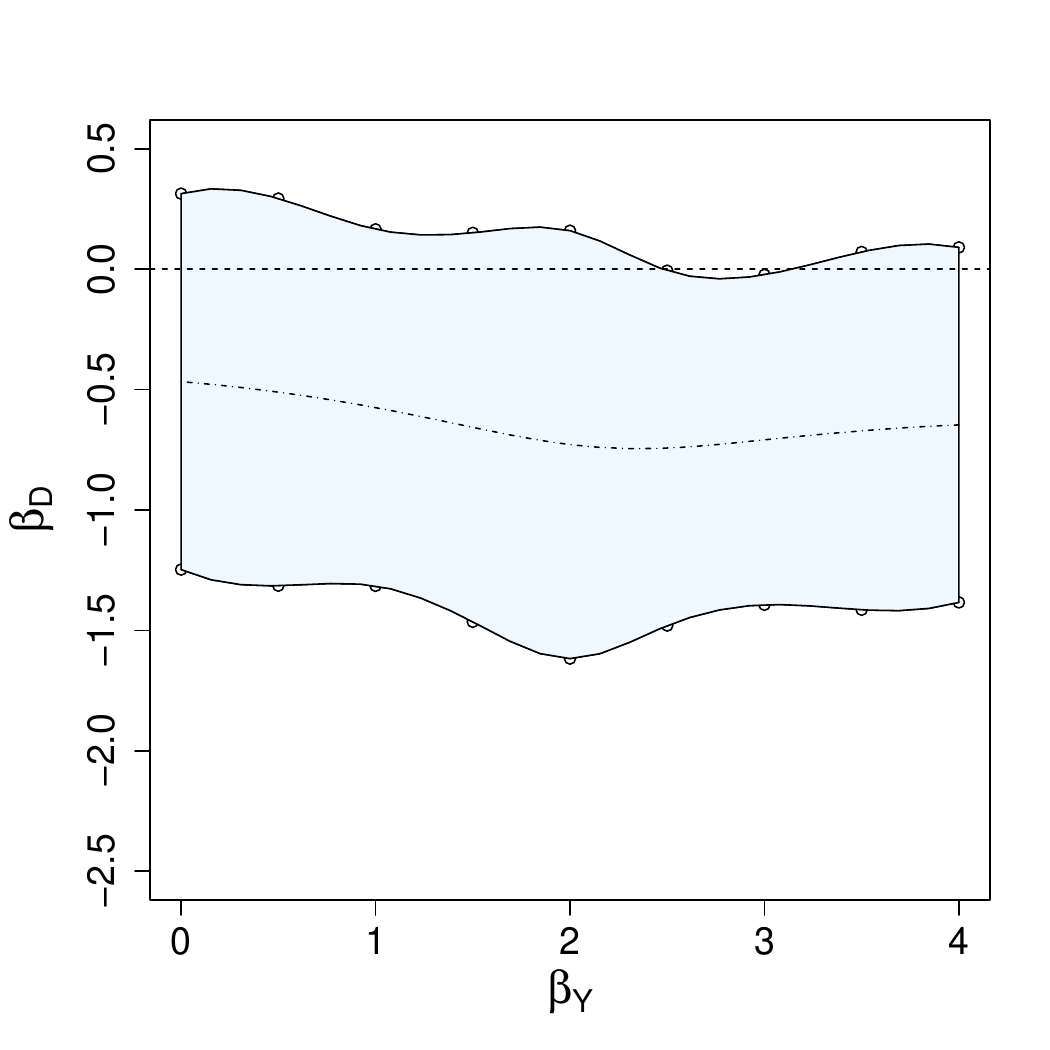}
\caption{\it 95\% confidence intervals of ATE (left) and $\beta_D$ (right) for given values of $\beta_Y$; 
the dashed lines denote point estimates.}
\label{fig:atetimeindsensi}
\end{figure}

\newpage
Figure \ref{fig:atetimeindsensi} shows the sensitivity analysis results. Note that as $\beta_Y $ increases past 1.3, $\alpha_D$ becomes non-significantly different from zero at 0.05 level two-sided. This makes sense because as more missing major birth defect outcomes become `yes', the rates of major birth defect between the exposed and unexposed groups become less differentiated. Meanwhile $\beta_D$ becomes more significantly less than zero in general, implying that exposure to the drug reduces the risk of spontaneous abortion. Such a mechanism allows more malformed fetus in the exposed group to develop into live born infants, which would have had a high chance of being spontaneously aborted had the women not been exposed to the medication. 

As discussed earlier $\boldV_{miss} = \emptyset$ in Assumption \ref{assump:mar} is somewhat strong  since we are not conditioning on any covariates once the SAB information has been accounted for. 
For pattern mixture model \citep{little2019statistical} sensitivity analysis has been proposed  by including a missing covariate in the outcome model with a sensitivity parameter \citep{leacy:etal, leurent:etal}. 
However, that is not applicable here with the selection model we are using. We are also reluctant to use inverse probability of missingness weighting as we would be multiplying three sets of weights 
leading to high variability and uninformative conclusions.

\begin{figure}
    \centering
    \includegraphics[scale = 0.29]{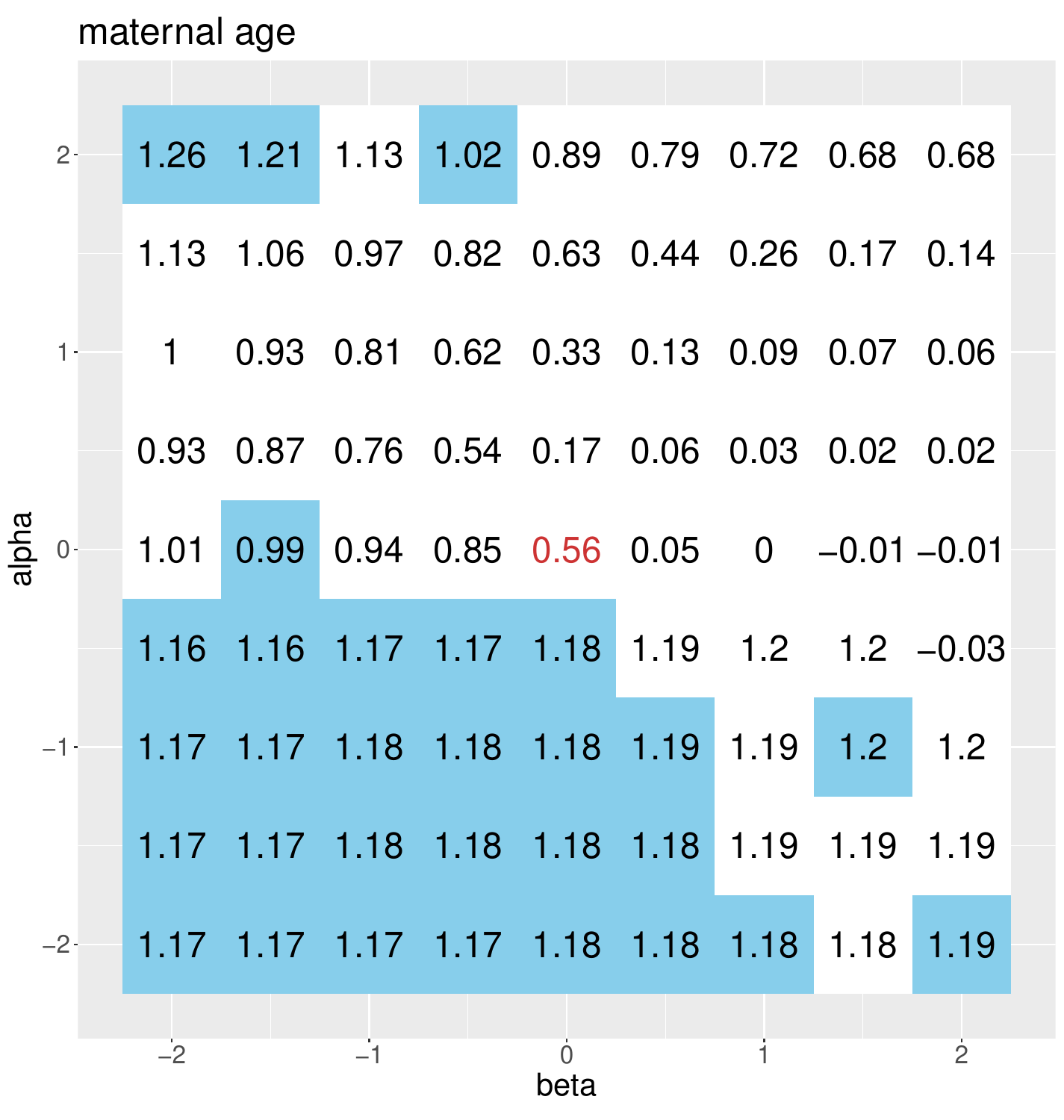}
    \includegraphics[scale = 0.29]{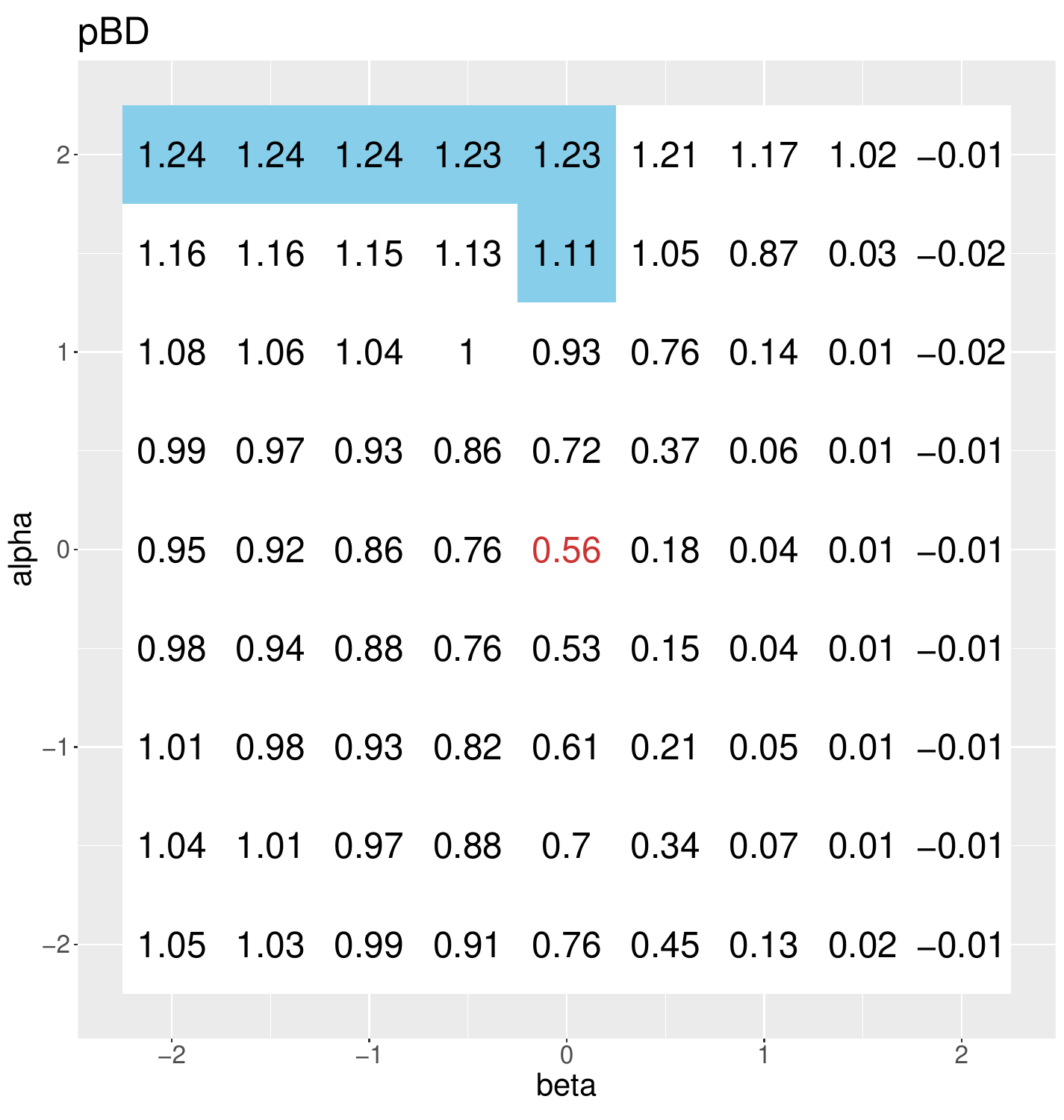}
    \includegraphics[scale = 0.29]{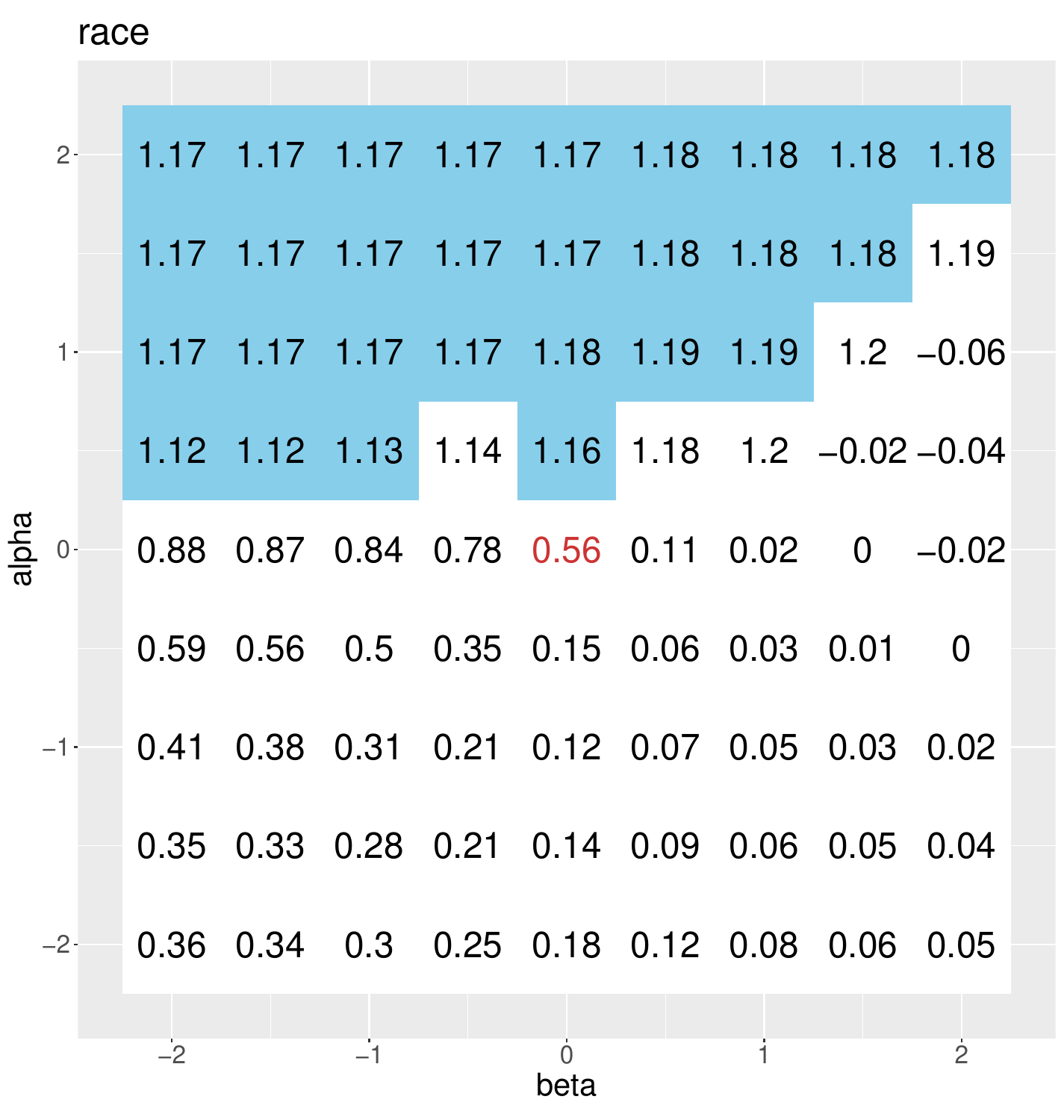}
    \includegraphics[scale = 0.29]{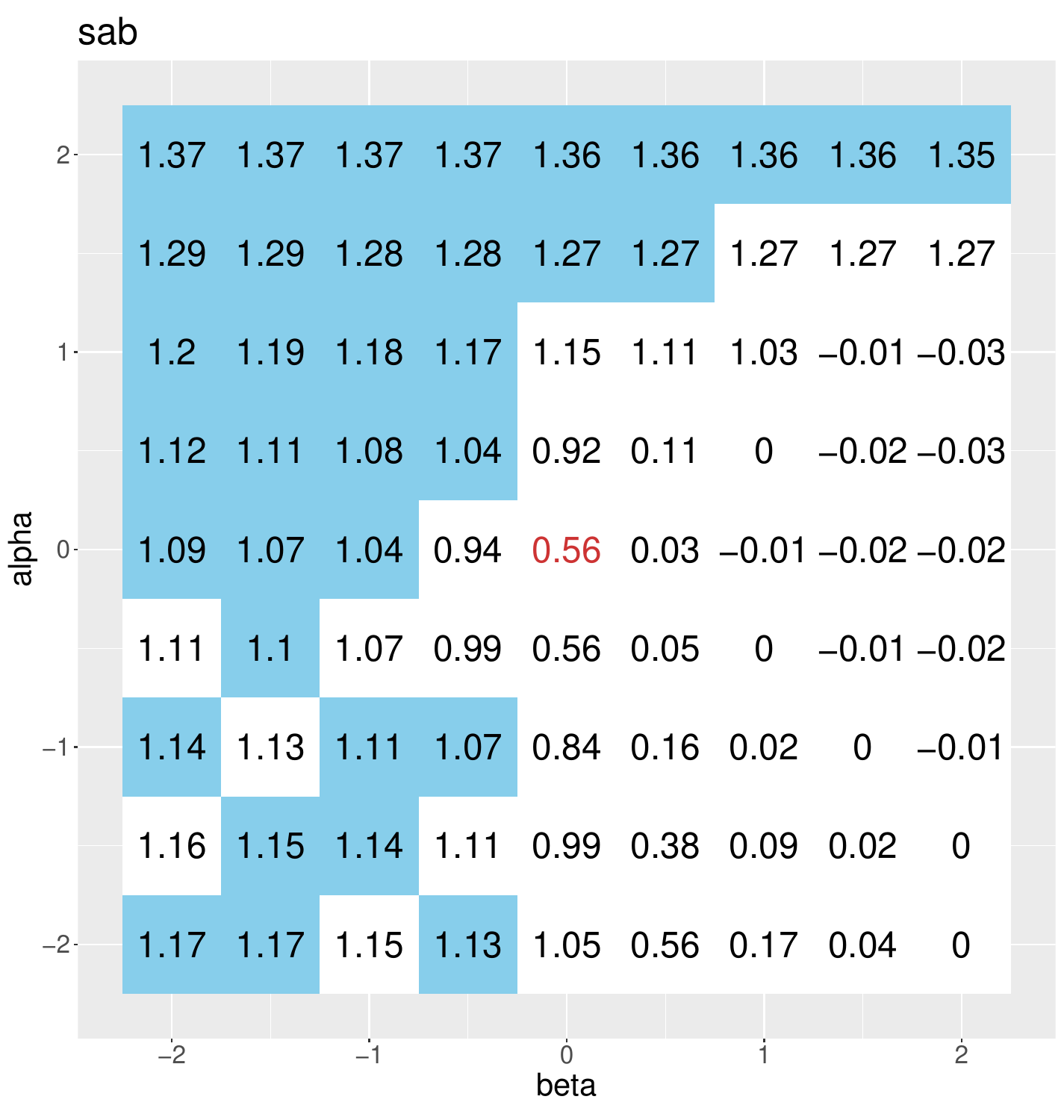}
    \includegraphics[scale = 0.29]{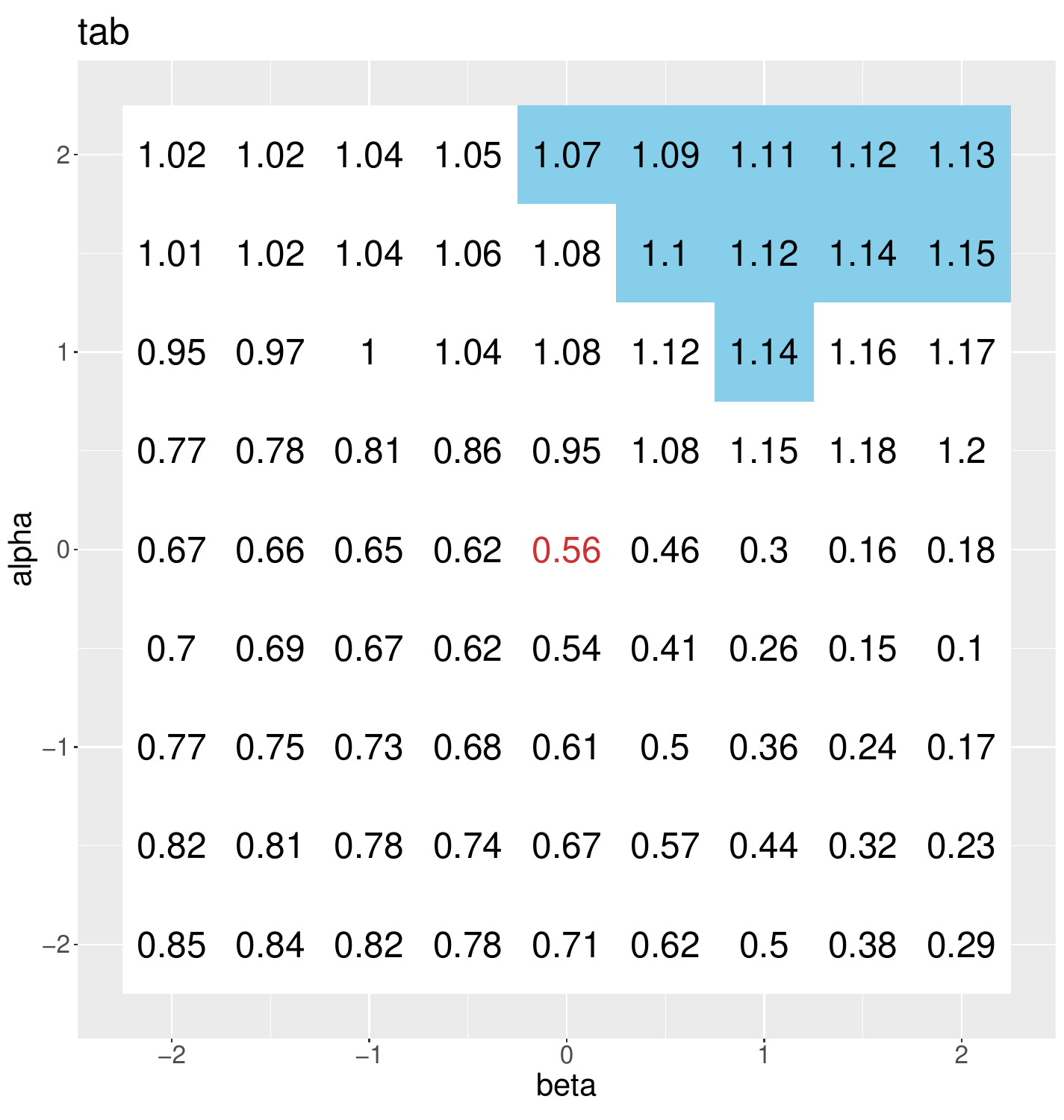}
    \caption{Sensitivity analysis for Assumption \ref{assump:mar}. Numbers in the cells represent point estimates of $\alpha_D$ and blue cells indicate significance at 0.05 level two-sided.  }
    \label{fig:sensimar}
\end{figure}

Based on expert knowledge, the potential predictors of missingness are previous SAB (yes/no), previous TAB (yes/no), maternal age ($\geq/<$ 35 years, maternal race/ethnicity (Others/non-Hispanic white), and history of a child with birth defects (yes/no). This is because these are predictors of SAB and LTFU, which are proxies of missing birth defect outcomes in this data set.
In the following we include one of these covariates at a time as $V_{miss}$,
additively into both \eqref{eq:yatetimeindcausalmodeliptw} and \eqref{eq:tatetimeindcausalmodeliptw}, and offset their parameters ($\alpha_{sensi}, \beta_{sensi}$). 
We remind the reader that by using the selection model to handle missing data, $V_{miss}$ is include in the  models \eqref{eq:yatetimeindcausalmodeliptw} and \eqref{eq:tatetimeindcausalmodeliptw}. We conduct our sensitivity analysis by running ($\alpha_{sensi}, \beta_{sensi}$) both between -2 and 2, with increment of 0.5. The results are shown in Figure \ref{fig:sensimar}. 

We note that when $\alpha_{sensi} \neq 0$, the interpretation of $\alpha_D$ in \eqref{eq:yatetimeindcausalmodeliptw} is no longer the ATE as described before. Instead it resembles a conditional average treatment effect (CATE) given the level of $\boldV_{miss}$, albeit with the sensitively parameter $\alpha_{sensi}$ in front of it. It is therefore difficult to interpret, and we suggest to focus on the sensitivity results with $\alpha_{sensi} = 0$. This also makes sense since, as mentioned earlier, the $\boldV_{miss}$ variables are mostly predictors of SAB for this data set. It is then seen that our conclusion of non-significant causal effect of the medication on birth defect holds unless, the effect of maternal age on SAB/stillbirth in model \eqref{eq:tatetimeindcausalmodeliptw} is as strong as -1.5 in terms of log relative risk, or the effect of previous SAB in the same model  is as strong as -1.0. Note that these indicate that older maternal age or previous SAB actually lowers the risk of SAB,  after adjusting for exposure and birth defect in \eqref{eq:tatetimeindcausalmodeliptw}. We suspect that these are unlikely scenarios in practice. 

\newpage
\subsection{Results on Principal Effects}\label{sec:peresult}

From Table \ref{table:observedgroup} we see that very few observed data points are available to estimate the parameters in the principal strata NN. Meanwhile 
most missing birth defect outcomes are from the observed groups $O(0, 1)$ and $O(1, 1)$, which consist of the latent principal strata (NS, NN) and NN, respectively. 
We have therefore decided to 
 focus on the estimation of the parameters in the SS strata, which appears to have most observed data points. 

The point estimates of the parameters and their standard errors (SE) etc., are presented in Table \ref{table:petimeindresult}. The results manifest that the drug has a significant effect in increasing SAB/Stillbirth, with a causal odds ratio close to 4. In addition, exposure to the drug has a significant positive effect on major birth defects. Note that this result, compared to that in Section \ref{sec:ateresult}, has a much narrower confidence interval, and thus we do not conduct a sensitivity analysis here.
 
\begin{table}
\centering
\caption{Parameter estimates from the OTIS data for PE analysis}
\label{table:petimeindresult}
\begin{tabular}{lcccc}
\hline
& Estimate (SE) & exp(Estimate) & 95\% CI of OR/HR  & $p$-value \\ \hline
$\alpha_{0, \SS} $ &-3.483  (0.2826) & 0.0307 & (0.0176, 0.0534) & $<$0.001 \\ 
$\alpha_{D, \SS} $ & 1.387  (1.738) & 4.002 & (0.1327, 120.7) & 0.425 \\ 
$\log(\mbox{OR}_M)$ &-0.3546(0.2961)&0.7014&(0.3926, 1.253)& 0.231\\   \hline
\end{tabular}
\end{table}

\section{Discussion}\label{sec:dis}

In this paper we have considered prospective pregnancy cohort studies where spontaneous abortion often results in unknown major birth defect outcomes. By convention of coding in the database a pregnancy is recorded as no birth defects unless one is found. Meanwhile it is well established that a malformed fetus has an increased chance of being  aborted \citep{warkany1978terathansaia}. By modeling the data mechanism using the terathanasia theory, we are able to turn the MNAR problem of major birth defects into an MAR setting by including information on the spontaneous abortion outcome. Our approach using \eqref{eq:tatetimeindcausalmodeliptw} is a selection model approach for handling informative missing \citep{little2019statistical}.
We note that missing outcomes have been considered in the causal inference literature; see for example \cite{frangakis1999, mealli:04}.

A second part of our work aims to properly handle the post-exposure variable live birth versus spontaneous abortion or stillbirth. We found the principal strata to be a useful framework for this. Due to the limited sample size and number of events, we are only able to estimate the principal effect in the largest stratum, i.e.~always survivors. A reviewer reminded us that the average treatment effect  can generally be derived as a weighted average of the principal strata effects, although in our case we are not able to   estimate all the principal effects. We also note that for non-linear models this would not be a simple or straightforward average. The approach, as well as R program, can be applied to larger data sets, where we should be able to estimate all principal effects. For other exposures that might not be protective against spontaneous abortion, one may need to consider four instead of three principal strata when the monotonicity assumption fails. 

As mentioned earlier while our data structure has some similarity with truncation by death in the principal strata setting, the latter typically concerns longitudinal outcomes over time. A common example is quality of life (QOL, or cognition function, etc.) that is truncated by death. In these settings the QOL outcome does not exist once a person dies, hence the only stratum of interest is always-survivors (and the principal effect is called survivor average causal effect or SACE). 
We note that principal stratification has also been considered in continuous time, including for semi-competing risks \citep{comment:2019, mattei:2020}. 
We also note the interesting work by \cite{elliott:etal} in developmental toxicology, who considered animal models where post-exposure litter size might have impact on outcomes such as birth weight, which were only defined for pups that were alive. They defined principal strata (PS) by the vector of pup survival statuses at each potential dose level, but otherwise  their PS definition was similar to ours. 
Birth defects, on the other hand, already exist before the end of the pregnancy, be it live birth or SAB/stillbirth. And as we have explained above, we are interested in the principal effects in all strata. 

In our analysis the major birth defect $Y$ is the outcome of interest. 
From Figure \ref{fig:dag}, however, $Y$ may also be seen as a mediator for the spontaneous abortion outcome $(M, T)$. The decomposition of the total effect of  exposure $D$ on SAB  into a direct effect and an indirect effect through $Y$ might be of interest, in understanding whether $D$ directly causes SAB or through major birth defects (and terathanasia). In this case the natural direct and the natural indirect effects might be suitable as they describe the underlying mechanism by which the exposure operates \citep{pearl2001direct}.

Finally, to estimate the average treatment effect (and similarly the principal effects), the structural logistic regression model \eqref{eq:yatetimeindcausalmodeliptw} for the causal odds ratio is saturated. However, we do impose a proportional hazards model in \eqref{eq:tatetimeindcausalmodeliptw}, as well as a propensity score model to generate inverse probability weights. Since the models might be wrong, it would be of interest to develop doubly robust or other robust approaches. This is an area for future work.

\acks{The authors gratefully acknowledge Professor Donald B.~Rubin for stimulating discussions on the principal effects. The authors also thank Ms.~Yuyao Wang for discussion regarding Assumption \ref{assump:quasitrunc}. 
AY was partially supported by the Achievement Rewards for College Scientists (ARCS) Scholarship.  AY was awarded the Student Poster Award based on this work at the 2019 Conference on Lifetime Data Science: Foundations and Frontiers. AY was awarded the 2020 Lifetime Data Science Section
of the American Statistical Association
Student Paper Award at the 2020 Joint Statistical Meetings. 
}


\newpage

\appendix

\section{Additional Data Summary}
\begin{table}[H]
\caption{Distribution of the maternal characteristics in OTIS study: mean (SD) or $n$ (\%).  }
\label{tab:char}
$$\begin{tabular}{lcc}
  \hline
Variables & Exposed ($n=336$) & Unexposed ($ n=158$) \\ 
  \hline
   Maternal Age&32.73(4.86)&33.44(5.09)\\
  Maternal Race&&\\
~~~Black  &12(3.6\%) &3(1.9\%) \\ 
~~~Hispanic  &35(10.4\%) &12(7.6\%) \\ 
~~~Not White or Black or Hispanic &20(5.9\%) &10(6.3\%)\\
~~~White Non-Hispanic &269 (80.0\%)&133(84.2\%)\\
Maternal Education &&\\
~~~$<$12 years &5(1.5\%) &4(2.5\%)\\
~~~12-15 years &108(32.1\%) &40(25.3\%)\\
~~~$>$15 years &223(66.4\%) &114(72.2\%)\\
SES &&\\
~~~High &310(92.8\%)&149(94.3\%)\\
~~~Low &24(7.2\%)&9(5.7\%)\\
BMI &&\\
 ~~~$<$18.5 (underweight)    &11(3.3\%)&7(4.4\%)\\
 ~~~$\geq$30 (obese) &68(20.2\%)&25(15.8\%)\\
 ~~~18.5-24.9 (normal weight)  &173(51.5\%)&98(62.0\%)\\
 ~~~  25-29.9 (overweight) &84(25.0\%)&28(17.8\%)\\
Gravidity   &&\\
 ~~~$>$1 &207 (61.6\%)&97(61.4\%)\\
 ~~~$=$1 &129(38.4\%) &61(38.6\%)\\
Parity &&\\
 ~~~$>$0 &156(46.4\%) &71(45.0\%)\\
 ~~~$=$0 &180(53.6\%) &87(55.0\%)\\
Previous SAB &&\\
 ~~~$>$0 &95(28.3\%)&45(28.5\%)\\
 ~~~$=$0 &241(71.7\%) &113(71.5\%)\\
Intended Pregnancy  &&\\
~~~No &96(28.6\%) &37(23.4\%)\\
~~~Yes &240(71.4\%) &121(76.6\%)\\
    Multiple Births&&\\
~~~No &323 (96.1\%)&156(98.7\%)\\
~~~Yes &13(3.9\%) &2(1.3\%)\\

    \hline
\end{tabular}$$

\end{table}

\begin{table}[H]
$$\begin{tabular}{lcc}
  \hline
  & Exposed ($n=336$) & Unexposed ($ n=158$) \\ 
  \hline
Previous TAB &&\\
~~~$>$0 &48(14.3\%) & 14(8.9\%)\\
~~~= 0 & 288(85.7\%) &144(91.1\%)\\
IVF &&\\
~~~No &312(93.1\%) &153(96.8\%)\\
~~~Yes &23(6.9\%) &5(3.2\%)\\
Vitamin and Folic Acid (Began prior to&&\\ conception, Post-conception only, &&\\
Have not taken at all)&&\\
~~~Began prior to conception     &199(59.2\%) &110(69.6\%)\\
~~~Have not taken at all      &2(0.6\%) &0(0\%)\\
~~~Post-conception only &135(40.2\%) &48(30.4\%)\\
 Alcohol  in the 1st Trimester  &&\\
~~~No &192(57.1\%) &88(55.7\%)\\
~~~Yes &144(42.9\%) &70(44.3\%)\\
Alcohol  between the Date of Conception&&\\ and Gestatinal Age of 20 Weeks &&\\
~~~No &190(56.5\%) &88(55.7\%)\\
~~~Yes &146(43.5\%) &70(44.3\%)\\
Tobacco  in the 1st Trimester  &&\\
~~~No &299(89.0\%) &141(89.2\%)\\
~~~Yes &37(11.0\%) &17(10.8\%)\\
Tobacco  between the Date of &&\\Conception and Gestatinal Age of 20 Weeks  &&\\
~~~No &301(89.6\%) &142(89.9\%)\\
~~~Yes &35(10.4\%) &16(10.1\%)\\
Other Major Known or Suspected &&\\ Human Teratogens  in the 1st Trimester    &&\\
    ~~~No &325(96.7\%) &147(93.0\%)\\
~~~Yes &11(3.3\%) &11(7.0\%)\\
Other Major Known or Suspected  Human &&\\Teratogens between the Date of &&\\Conception and Gestatinal Age of 20 Weeks &&\\
    ~~~No &323(96.1\%) &147(93.0\%)\\
~~~Yes &13(3.9\%) &11(7.0\%)\\
History of a Child with Birth Defects  &&\\
    ~~~No &321(95.5\%) &150(94.9\%)\\
~~~Yes &15(4.5\%) &8(5.1\%)\\
    \hline
\end{tabular}$$

\end{table}

\begin{table}[H]
$$\begin{tabular}{lcc}
  \hline
  & Exposed ($n=336$) & Unexposed ($ n=158$) \\ 
  \hline
Antidepressant in the 1st Trimester    &&\\
    ~~~No &312(92.9\%) &144(91.1\%)\\
~~~Yes &24(7.1\%) &14(8.9\%)\\
Antidepressant between the Date of&&\\ Conception and Gestatinal Age of 20 Weeks &&\\
    ~~~No &322(95.8\%) &152(96.2\%)\\
~~~Yes &14(4.2\%) &6(3.8\%)\\
Pre-gestational Hypertension&&\\
    ~~~No &317(94.3\%) &148(93.7\%)\\
~~~Yes &19(5.7\%) &10(6.3\%)\\
Thyroid Disease  &&\\
       ~~~No &292(86.7\%) &135(85.4\%)\\
~~~Yes &44(13.3\%) &23(4.6\%)\\
Other Psychiatric Conditions &&\\
        ~~~No &277(86.9\%) &128(81.0\%)\\
~~~Yes &59(13.1\%) &30(19.0\%)\\
Infections in the 1st Trimester &&\\
    ~~~No &219(65.6\%) &97(61.4\%)\\
~~~Yes &115(34.4\%) &61(38.6\%)\\
Infection  between the Date of Conception &&\\
and Gestatinal Age of 20 Weeks &&\\
        ~~~No &336(100\%) &158(100\%)\\
~~~Yes &0(0\%) &0(0\%)\\
Primary Disease &&\\
 ~~~Psoriasis &53(15.8\%) &46(29.1\%)\\
 ~~~RA, JRA, PsA or AS &283(84.2\%) &112(70.9\%)\\
Other Autoimmune Disease &&\\
    ~~~No &295(87.8\%) &113(89.9\%)\\
~~~Yes &41(12.2\%) &45(10.1\%)\\
No. of Autoimmune Diseases &&\\
 ~~~$\leq$ 2 &321(95.5\%) &153(96.8\%)\\
 ~~~$>$ 2 &15(4.5\%) &5(3.2\%)\\
     \hline
\end{tabular}$$

\end{table}

\begin{table}[H]
$$\begin{tabular}{lcc}
  \hline
  & Exposed ($n=336$) & Unexposed ($ n=158$) \\ 
  \hline
Prednisone and/or Systemic Oral&&\\ Corticosteroid in the 1st Trimester  &&\\
     ~~~No &225(67.2\%) &113(71.5\%)\\
~~~Yes &110(32.8\%) &45(28.5\%)\\
Prednisone and/or Systemic Oral &&\\Corticosteroid between the Date of Conception&&\\ and Gestatinal Age of 20 Weeks  &&\\
~~~No 				&240(71.6\%) 	&117(74.1\%)\\
~~~Yes 				&95(28.4\%) 		&41(25.9\%)\\
Prednisone and/or Systemic Oral Corticosteroid &&\\Average Dose in the 1st Trimester &22.7(53.7)&18.5(51.6)\\
Prednisone and/or Systemic Oral&&\\ Corticosteroid Average Dose between the Date of&&\\ Conception and Gestatinal Age of 20 Weeks &3.6(10.3) &2.4(6.7)\\
Prednisone and/or Systemic Oral Corticosteroid&&\\
 Duration in the 1st Trimester  &&\\
~~~0-4.0 Weeks  &37(11.1\%) &12(7.6\%)\\
~~~4.1-6.0 Weeks &10(3.0\%) &3(1.9\%)\\
~~~6.1-12.0 Weeks  &62(18.6\%) &30(19.0\%)\\
 ~~~No Exp in 1st  &225(67.4\%) &113(71.5\%)\\
Prednisone and/or Systemic Oral &&\\
Corticosteroid Duration between the Date of &&\\
Conception and Gestatinal Age of 20 Weeks &&\\
~~~$>$12 Weeks    &33(9.9\%) &15(9.6\%)\\
~~~0-4.0 Weeks   &33(9.9\%) &12(7.6\%)\\
~~~4.1-6.0 Weeks &6(1.8\%) &6(3.8\%)\\        
~~~6.1-12.0 Weeks &23(6.9\%) &7(4.5\%)\\
~~~No Exp prior to 18 WeeksPC &240(71.6\%) &117(74.5\%)\\
Years Since Diagnosis of Primary Disease&9.72(7.96) &10.20(9.20)\\
Disease Severity Score at Intake  - RA1 &0.412(0.532) &0.466(0.616) \\
 Disease Severity Score at Intake  - RA2 &25.0(27.9) &23.9(27.14) \\
Disease Severity Score at Intake  - RA3 &21.0(24.8) &20.1(24.23) \\
  Disease Severity Score at Intake  - PsO1 &0.53(1.17) &0.62(1.08) \\
  Disease Severity Score at Intake  - PsO2 &22.9(37.9) &27.5(40.42) \\
  Country of Residence (U.S., Canada)   & &\\
  ~~~Canada & 28 (8.4\%) & 25 (15.9\%) \\
 ~~~ U.S.& 307 (91.6\%) & 132 (84.1\%) \\
  Pregnancy Weight (kg)  & 70.5 (18.18) & 68.5 (16.65) \\
   \hline
\end{tabular}$$

\end{table}

\section{Additional Material of Section \ref{sec:ate}}
\subsection{Identification}
We justify that the weights we apply create a pseudopopulation where the treatment is randomized and there is no truncation. Let $f(\cO) = f(V, D, Y, X, \Delta, O, Q)$ loosely denote the density ignoring truncation, then the actual distribution $\cQ$ we are sampling from is
\eqn
d\cQ = f(V, D, Y, X, \Delta, O|X > Q)d\cO.
\ee
The distribution $\cP$ we target, in which the treatment is randomized and there is no truncation is,
\eqn
d\cP = f(V)f(D)f(Y, X, \Delta, O|D, V)d\cO.
\ee
Therefore, the weight that one shall adopt, is the Radon-Nikodym derivative, $d\cP/d\cQ$, which in this case is
\eqn
\frac{d\cP}{d\cQ} &=& \frac{f(Y, X, \Delta, O|D, V)f(D)f(V)}{f(V, D, Y, X, \Delta, O, X > Q)/\P(X > Q)}\\
&=&\frac{f(Y, X, \Delta, O|D, V)f(D)f(V)}{f(V)f(D|V)f(Y, X, \Delta, O|D, V)\int_0^X f(Q|Y, X, \Delta, O, D, V)dQ/\P(X > Q)}\\
&=&\frac{f(Y, X, \Delta, O|D, V)f(D)f(V)}{f(V)f(D|V)f(Y, X, \Delta, O|D, V)\int_0^X f(Q|D, V)dQ/\P(X > Q)}\\
&=&\frac{f(D)\P(X > Q)}{f(D|V)P(Q < q|D, V)|_{q = X}}.
\ee
Our weights differ by a constant. Note that by using potential outcomes
\eqn
d\cP = f(V)f(D)f(Y(D), X(D), \Delta(D), O(D)|V)d\cO,
\ee
which by integrating $V$, conditional randomization and consistency marginalizes to
\eqn
d\cP = f(D)f(Y(D), X(D), \Delta(D), O(D))d\cO,
\ee
We first write down the complete data likelihood that augments the counterfactual outcomes, missing outcomes, censoring, on a population level, which by independence assumptions factorizes into
\eqn
&&\P(Y(d), T(d), C(d), O(d)) \nonumber\\
&=&\P(Y(d))f_T(T_{obs}(d)|Y(d)))^{\Delta(d)} f_T(T_{mis}(d)|Y(d))^{1 - \Delta}\nonumber\\
&&~~\cdot f_C(C_{obs}(d)|Y(d), T(d))^{1 - \Delta(d)} f_C(C_{mis}(d)|Y(d), T(d))^{\Delta(d)} \P(O(d)|T(d), C(d), Y_{obs}(d), Y_{mis}(d))\nonumber\\
&=&\P(Y(d))f_T(T_{obs}(d)|Y(d)))^{\Delta(d)} f_T(T_{mis}(d)|Y(d))^{1 - \Delta(d)}\nonumber\\
&&~~\cdot f_C(C_{obs}(d)|Y(d))^{1 - \Delta(d)} f_C(C_{mis}(d)|Y(d))^{\Delta(d)} \P(O(d)|X(d), \Delta(d), Y_{obs}(d)).
\ee
By integrating the complete likelihood with respect to $Y_{obs}(d)$, $X(d)$, $\Delta(d)$, $O(d)$, we arrive at
\eqn
&&\P(Y_{obs}(d))f_T(X(d)|Y_{obs}(d)))^{\Delta(d)} S(X(d)|Y_{obs}(d))^{1 - \Delta(d)}/\P(T(d) > Q(d))\\
&&~~\cdot f_C(X(d)|Y_{obs}(d))^{1 - \Delta(d)} S_C(X(d)|Y_{obs}(d))^{\Delta(d)} \P(O(d)|X(d), \Delta(d), Y_{obs}(d)),
\ee
which implies that parameters of interest ``ignore'' the distribution $f(C(d)|Y_{obs}(d))$, $\P(O(d)|Y_{obs}(d), X(d), \Delta(d))$. Therefore we can directly maximize $\P(Y_{obs}(d))f_T(X(d)|Y_{obs}(d)))^{\Delta(d)} S(X(d)|Y_{obs}(d))^{1 - \Delta(d)}/\P(T(d) > Q(d))$.

\subsection{E-step functions}
At the $(t + 1)$-th step $(t = 0,1,\cdots)$, we have
\eqn
\P^{(t)}(Y_i^{mis} = 1|D_i, T_i > X_i, Q_i) = \frac{\pi_i^{(t)}S_i^{(t)}(X_i|Y_i = 1)}{\pi_i^{(t)}S_i^{(t)}(X_i|Y_i = 1) + (1 - \pi_i^{(t)})S_i^{(t)}(X_i|Y_i = 0)},
\ee
\eqn
\P^{(t)}(Y_i^{mis} = 1|D_i, T_i = X_i, Q_i) = \frac{\pi_i^{(t)}f_i^{(t)}(X_i|Y_i = 1)}{\pi_i^{(t)}f_i^{(t)}(X_i|Y_i = 1) + (1 - \pi_i^{(t)})f_i^{(t)}(X_i|Y_i = 0)},
\ee
\eqn
\P^{(t)}(T_{i1} = t_k|D_i, Y_i = 1, Q_i) = \frac{\mathbbm{1}(t_k < Q_i)f_i^{(t)}(t_k|Y_i = 1)}{1 - S_i^{(t)}(Q_i|Y_i = 1)},
\ee
\eqn
\P^{(t)}(T_{i1} = t_k|D_i, Y_i = 0, Q_i) = \frac{\mathbbm{1}(t_k < Q_i)f_i^{(t)}(t_k|Y_i = 0)}{1 - S_i^{(t)}(Q_i|Y_i = 0)},
\ee
In the above the first two expressions are for missing outcomes, and the rest for handling left truncation. With a more granular classification of the subjects, the Q-function \eqref{eq:ateqfunc} becomes
\eqnn
&&Q(\btheta|\btheta^{(t)})\\
&=& \sum_{i:\Delta_i = 0, O_i = 1} w_i \Big\{Y_i\log \pi_i + (1 - Y_i)\log(1 - \pi_i) + \log S_i(X_i)\Big\}\\
&+& \sum_{i:\Delta_i = 0, O_i = 0}w_i \Big\{\P_i^{(t)}(Y_i^{mis} = 1)\log \pi_i + \P_i^{(t)}(Y_i^{mis} = 0)\log(1 - \pi_i) \\
&&+  \P_i^{(t)}(Y_i^{mis} = 1)\log S_i(X_i|Y_i = 1) +  \P_i^{(t)}(Y_i^{mis} = 0)\log S_i(X_i|Y_i = 0)\Big\}\\
&+& \sum_{i:\Delta_i = 1, O_i = 1}w_i \Big\{Y_i\log \pi_i + (1 - Y_i)\log(1 - \pi_i) + \log f_i(X_i)\Big\}\\
&+& \sum_{i:\Delta_i = 1, O_i = 0} w_i \Big\{\P_i^{(t)}(Y_i^{mis} = 1)\log \pi_i + \P_i^{(t)}(Y_i^{mis} = 0)\log(1 - \pi_i) \\
&&+ \P_i^{(t)}(Y_i^{mis} = 1)\log f_i(X_i|Y_i = 1) + \P_i^{(t)}(Y_i^{mis} = 0)\log f_i(X_i|Y_i = 0)\Big\}.
\een

\subsection{Simulation details}

We set $V = \emptyset$. We initiate $(D, Y, X = min(T_0, C), \Delta = \mathbbm{1}(T \leq C), O)$ by 
\begin{equation}
P(D = 1) = 0.5,
\end{equation}
\begin{equation}
P(Q(d) > t) = \exp(-(0.5 + 0.1d)t),
\end{equation}
\begin{equation}
P(Y(d) = 1) = \frac{1}{1 + \exp(1 -0.3d)},
\end{equation}
\begin{equation}
P(T(d) > t|Y(d) = y) = \exp\{-0.0375t\exp(0.2  d + 0.6 y)\} \mathbbm{1}(t <= 45),
\end{equation}
\begin{equation}
P(C(d) > t + 10|Q(d)) = \exp(-0.005t) \mathbbm{1}(t <= 30),
\end{equation}
\begin{equation}
P(O(d) = 1|Y(d), T(d), M(d))  = \frac{1}{1 + \exp(-3 + 0.1 \log(X(d)))},
\end{equation}
and we only keep those with $T_i > Q_i$  and those $Y_i$ when $O_i = 1$ to reflect left truncation and missing outcomes.  We consider sample size $N = 500$. We set an administrative censoring at $t = 40$ and repeat time $B = 100$ to attain a bootstrapped variance for bootstrapped normal confidence intervals.

\section{Additional Materials of Section \ref{sec:pe}}

\subsection{Identification}
As in the ATE case, we justify that the weights we apply create a pseudopopulation where the treatment is randomized and there is no truncation. The main difference is that we need to leverage $M$. Let $f(\cO) = f(V, D, Y, X, \Delta, O, Q, M)$ loosely denote the density ignoring truncation, then the actual distribution $\cQ$ we are sampling from is
\eqn
d\cQ = f(V, D, Y, X, \Delta, O, M|X > Q)d\cO.
\ee
The distribution $\cP$ we target, in which the treatment is randomized and there is no truncation is,
\eqn
d\cP = f(V)f(D)f(Y, X, \Delta, O, M|D, V)d\cO.
\ee
Therefore, the weight that one shall adopt, is the Radon-Nikodym derivative, $d\cP/d\cQ$, which in this case is
\eqn
\frac{d\cP}{d\cQ} &=& \frac{f(Y, X, \Delta, O, M|D, V)f(D)f(V)}{f(V, D, Y, X, \Delta, O, M, X > Q)/\P(X > Q)}\nonumber\\
&=&\frac{f(Y, X, \Delta, O, M|D, V)f(D)f(V)}{f(V)f(D|V)f(Y, X, \Delta, O, M|D, V)\int_0^X f(Q|Y, X, \Delta, O, D, V, M)dQ/\P(X > Q)}\nonumber\\
&=&\frac{f(Y, X, \Delta, O, M|D, V)f(D)f(V)}{f(V)f(D|V)f(Y, X, \Delta, O, M|D, V)\int_0^X f(Q|D, V)dQ/\P(X > Q)}\nonumber\\
&=&\frac{f(D)\P(X > Q)}{f(D|V)P(Q < q|D, V)|_{q = X}}.
\ee
Therefore by the same reasoning as in ATE, we can use the same weights to handle nonrandomized treatment allocation and left truncation.

\section{Birth defect data analysis}

When conducting the sensitivity analysis for this data set, we offset some of the parameters. For example, in Section \ref{sec:ateresult} we offset $\beta_Y$ at some pre-specified value $\beta_{Y, 0}$. To obtain $\hat \btheta$, it suffices to set $\beta_Y^{(t)} = \beta_{Y, 0}$ in step $t$ of the ES algorithm. All expressions are modified to accommodate such changes. 

\begin{figure}[H]
\centering
\includegraphics[scale=0.75]{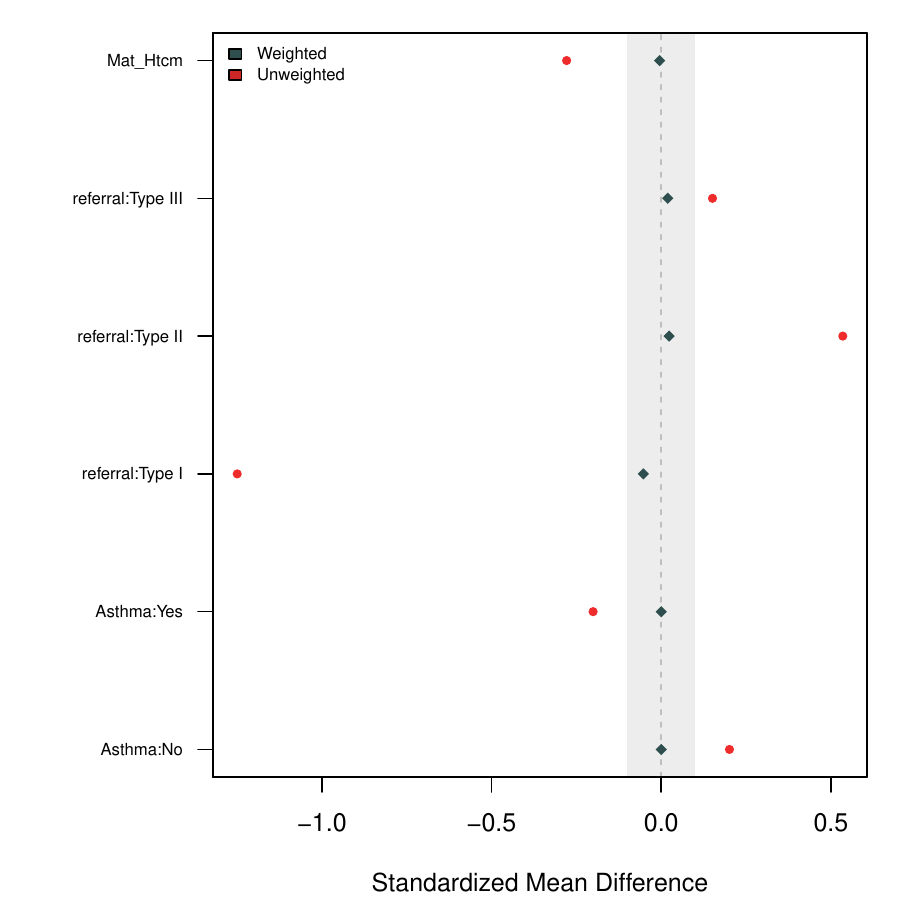}
\caption{Balance assessment via propensity score}
\label{fig:psbalancecheck}
\end{figure}

\newpage
Figure \ref{fig:psbalancecheck} shows the the standardized mean difference between the exposed  and the unexposed groups before and after weighting by the inverse of the propensity score.

\vskip 0.2in
\bibliography{phmm2_preg}

@string{JASA    = {Journal of the American Statistical Association}}

@string{Bka             = {Biometrika}}

@string{Bx              = {Biometrics}}

@string{Bcs              = {Biometrics}}

@string{lida             = {Lifetime Data Analysis}}

@string{CJS             = {Canadian Journal of Statistics}}

@string{aje    = {American Journal of Epidemiology}}

@article{bres:well,
	Author = {Breslow, Norman E and Wellner, Jon A},
	Journal = {Scandinavian Journal of Statistics},
	Number = {1},
	Pages = {86--102},
	Publisher = {Wiley Online Library},
	Title = {Weighted Likelihood for Semiparametric Models and Two-phase Stratified Samples, with Application to {C}ox Regression},
	Volume = {34},
	Year = {2007}}

@article{comment:2019,
	Author = {Comment, L and Mealli, F and Haneuse, S and Zigler, C},
	Journal = {arXiv:1902.09304},
	Title = {Survivor average causal effects for continuous time: a principal stratification approach to causal inference with semicompeting risks},
	Year = {2019}}

@article{d1998propensity,
  title={Propensity score methods for bias reduction in the comparison of a treatment to a non-randomized control group},
  author={D'Agostino Jr, Ralph B},
  journal={Statistics in Medicine},
  volume={17},
  number={19},
  pages={2265--2281},
  year={1998},
  publisher={Wiley Online Library}
}

@article{ding2011identifiability,
	Author = {Ding, Peng and Geng, Zhi and Yan, Wei and Zhou, Xiao-Hua},
	Journal = {Journal of the American Statistical Association},
	Number = {496},
	Pages = {1578--1591},
	Publisher = {Taylor \& Francis},
	Title = {Identifiability and estimation of causal effects by principal stratification with outcomes truncated by death},
	Volume = {106},
	Year = {2011}}

@article{ding2018causal,
	Author = {Ding, Peng and Li, Fan and others},
	Journal = {Statistical Science},
	Number = {2},
	Pages = {214--237},
	Publisher = {Institute of Mathematical Statistics},
	Title = {Causal inference: A missing data perspective},
	Volume = {33},
	Year = {2018}}

@article{elashoff2004algorithm,
	Author = {Elashoff, Michael and Ryan, Louise},
	Journal = {Journal of Computational and Graphical Statistics},
	Number = {1},
	Pages = {48--65},
	Publisher = {Taylor \& Francis},
	Title = {An EM algorithm for estimating equations},
	Volume = {13},
	Year = {2004}}

@article{elliott:etal,
	Author = {Elliott, M R and Joffe, M. M. and Chen, Z.},
	Journal = bcs,
	Number = {2},
	Pages = {352-360},
	Title = {A potential outcomes approach to developmental toxicity analyses},
	Volume = {62},
	Year = {2000}}

@book{faig2013,
	Address = {San Diego},
	Author = {Faig, W},
	Publisher = {Ph.D. thesis of the University of California},
	Title = {A Joint Marginal-Conditional For Multivariate Longitudinal Data and A Cure-Rate Model For Left-Truncated and Right-Censored Data},
	Year = {2013}}

@article{farewell,
	Author = {Farewell, V T},
	Journal = bx,
	Pages = {1041-1046},
	Title = {The use of mixture models for the analysis of survival data with long-term survivors},
	Volume = 38,
	Year = 1982}

@article{farewell:86,
	Author = {Farewell, V T},
	Journal = cjs,
	Pages = {257-262},
	Title = {Mixture models in survival analysis: Are they worth the risk?},
	Volume = 14,
	Year = 1986}

@article{frangakis1999,
	Author = {Frangakis, Constantine E and Rubin, Donald B},
	Journal = {Biometrika},
	Number = {2},
	Pages = {365-379},
	Title = {Addressing complications of intention-to-treat analysis in the combined presence of all-or-none treatment-noncomplicance and subsequent missing outcomes},
	Volume = {86},
	Year = {1999}}

@article{frangakis2002principal,
	Author = {Frangakis, Constantine E and Rubin, Donald B},
	Journal = {Biometrics},
	Number = {1},
	Pages = {21--29},
	Publisher = {Wiley Online Library},
	Title = {Principal stratification in causal inference},
	Volume = {58},
	Year = {2002}}

@article{frumento2012evaluating,
	Author = {Frumento, Paolo and Mealli, Fabrizia and Pacini, Barbara and Rubin, Donald B},
	Journal = {Journal of the American Statistical Association},
	Number = {498},
	Pages = {450--466},
	Publisher = {Taylor \& Francis},
	Title = {Evaluating the effect of training on wages in the presence of noncompliance, nonemployment, and missing outcome data},
	Volume = {107},
	Year = {2012}}

@article{holland1986statistics,
	Author = {Holland, Paul W},
	Journal = {Journal of the American Statistical Association},
	Number = {396},
	Pages = {945--960},
	Publisher = {Taylor \& Francis},
	Title = {Statistics and causal inference},
	Volume = {81},
	Year = {1986}}

@article{hou2018nonparametric,
	Author = {Hou, Jue and Chambers, Christina D and Xu, Ronghui},
	Journal = {Lifetime Data Analysis},
	Volume = {24},
	Number = {4},
	Pages = {612-651},
	Publisher = {Springer},
	Title = {A nonparametric maximum likelihood approach for survival data with observed cured subjects, left truncation and right-censoring},
	Year = {2018}}

@Manual{nloptr,
    title = {The NLopt nonlinear-optimization package},
    author = {Johnson, Steven G. },
    year={2021},
    url = {http://ab-initio.mit.edu/nlopt},
}

@book{kosorok,
	Author = {Kosorok, M. R.},
	Publisher = {Springer},
	Title = {Introduction to Empirical Processes and Semiparametric Inference},
	Year = 2008}

@article{kuk1992mixture,
  title={A mixture model combining logistic regression with proportional hazards regression},
  author={Kuk, Anthony YC and Chen, Chen-Hsin},
  journal={Biometrika},
  volume={79},
  number={3},
  pages={531--541},
  year={1992},
  publisher={Oxford University Press}
}

@article{leacy:etal,
	Author = { Leacy, F P and Floyd, S and  Yates, T A and  White, I R},
	Journal = aje,
	Pages = {304–315},
	Title = {Analyses of Sensitivity to the Missing-at-Random Assumption Using Multiple Imputation With Delta Adjustment: Application to a Tuberculosis/{HIV} Prevalence Survey With Incomplete HIV-Status Data},
	Volume = 185,
	Year = 2017}

@article{leurent:etal,
	Author = {Leurent, B and Gomes, M and Faria, R and Morris, S and Grieve, R and Carpenter, J R},
	Journal = {Pharmaco{E}conomics},
	Pages = {889-901},
	Title = {Sensitivity Analysis for Not-at-Random Missing Data in Trial-Based Cost-Effectiveness Analysis: A Tutorial},
	Volume = 36,
	Year = 2018}

@book{little2019statistical,
  title={Statistical {A}nalysis with {M}issing {D}ata},
  author={Little, Roderick JA and Rubin, Donald B},
  year={2019},
  publisher={John Wiley \& Sons}
}

@article{lu:ying,
	Author = {Lu, W and Ying, Z},
	Journal = Bka,
	Pages = {331-343},
	Title = {On semiparametric transformation cure models},
	Volume = 91,
	Year = 2004}

@article{mattei:2020,
	Author = {Mattei, A and Ding, P and Mealli, F},
	Journal = {arXiv:2002.11989},
	Title = {Assessing causal effects in the presence of treatment switching through principal stratification},
	Year = {2020}}

@article{mealli:04,
	Author = {Mealli, F and Imbens, G W and Ferro, S and Biggeri, A},
	Journal = {Biostatistics},
	Pages = {207-222},
	Title = {Analyzing a randomized trial on breast self{-}examination with noncomplicance and missing outcomes},
	Volume = 5,
	Year = {2004}}

@article{meister,
	Author = {Meister, R and Schaefer, C},
	Journal = {Reproductive Toxicology},
	Pages = {31-35},
	Title = {Statistical methods for estimating the probability of spontaneous abortion in observational studies -- analyzing pregnancies exposed to coumarin derivatives},
	Volume = 26,
	Year = 2008}

@article{neyman1923applications,
  title={Sur les applications de la th{\'e}orie des probabilit{\'e}s aux experiences agricoles: Essai des principes},
  author={Neyman, Jersey},
  journal={Roczniki Nauk Rolniczych},
  volume={10},
  pages={1--51},
  year={1923}
}

@inproceedings{pearl2001direct,
	Author = {Pearl, Judea},
	Booktitle = {Proceedings of the {S}eventeenth {C}onference on {U}ncertainty in {A}rtificial {I}ntelligence},
	Organization = {Morgan Kaufmann Publishers Inc.},
	Pages = {411--420},
	Title = {Direct and indirect effects},
	Year = {2001}}

@article{peng2000nonparametric,
  title={A nonparametric mixture model for cure rate estimation},
  author={Peng, Yingwei and Dear, Keith BG},
  journal={Biometrics},
  volume={56},
  number={1},
  pages={237--243},
  year={2000},
  publisher={Wiley Online Library}
}

@article{qin:etal:2011,
	Author = {Qin, J. and Ning, J and Liu, H and Shen, Y},
	Journal = jasa,
	Pages = {1434-1449},
	Title = {Maximum likelihood estimations and {EM} algorithms with length-biased data},
	Volume = {106},
	Year = {2011}}

@manual{twang,
	Author = {Ridgeway, Greg and McCaffrey, Dan and Morral, Andrew and Griffin, Beth Ann and Burgette, Lane},
	Note = {R package version 1.5},
	Title = {twang: Toolkit for Weighting and Analysis of Nonequivalent Groups},
	Url = {https://CRAN.R-project.org/package=twang},
	Year = {2017}}

@article{robins1992identifiability,
  title={Identifiability and exchangeability for direct and indirect effects},
  author={Robins, James M and Greenland, Sander},
  journal={Epidemiology},
  volume={3},
  number={2},
  pages={143--155},
  year={1992},
  publisher={JSTOR}
}

@article{rosen:etal,
	Author = {Rosen, O and Jiang, W and Tanner, M A},
	Journal = bka,
	Pages = {391-404},
	Title = {Mixtures of marginal models},
	Volume = {87},
	Year = {2000}}

@article{rosenbaum1984consequences,
  title={The consequences of adjustment for a concomitant variable that has been affected by the treatment},
  author={Rosenbaum, Paul R},
  journal={Journal of the Royal Statistical Society: Series A (General)},
  volume={147},
  number={5},
  pages={656--666},
  year={1984},
  publisher={Wiley Online Library}
}

@article{rosenbaum1983central,
  title={The central role of the propensity score in observational studies for causal effects},
  author={Rosenbaum, Paul R and Rubin, Donald B},
  journal={Biometrika},
  volume={70},
  number={1},
  pages={41--55},
  year={1983},
  publisher={Oxford University Press}
}

@article{rubin1974,
  title={Estimating causal effects of treatments in randomized and nonrandomized studies},
  author={Rubin, Donald B},
  journal={Journal of Educational Psychology},
  volume={66},
  number={5},
  pages={688-701},
  year={1974}
}

@article{runarsson2005search,
  title={Search biases in constrained evolutionary optimization},
  author={Runarsson, Thomas Philip and Yao, Xin},
  journal={IEEE Transactions on Systems, Man, and Cybernetics, Part C (Applications and Reviews)},
  volume={35},
  number={2},
  pages={233--243},
  year={2005},
  publisher={IEEE}
}

@article{ santiago:etal,
    Author = {Santiago, K Y and Porchia, L M and Lopez{-}Bayghen, E},
    Title = {Endometrial preparation with etanercept increased embryo
implantation and live birth rates in women suffering from recurrent
implantation failure during {IVF}},
    Journal = {Reproductive Biology},
    Year = 2021,
    Volume = {21},
    Pages = {http://dx.doi.org/10.1016/j.repbio.2021.100480}
    }

@article{sy:taylor,
	Author = {Sy, Judy P and Taylor, Jeremy MG},
	Journal = {Biometrics},
	Number = {1},
	Pages = {227--236},
	Publisher = {Wiley Online Library},
	Title = {Estimation in a Cox proportional hazards cure model},
	Volume = {56},
	Year = {2000}}

@article{vaid:xu:00,
	Author = {Vaida, F. and Xu, R.},
	Journal = {Statistics in Medicine},
	Pages = {3309--3324},
	Title = {Proportional hazards model with random effects},
	Volume = {19},
	Year = {2000}}

@article{vardi:89,
	Author = {Vardi, Y},
	Journal = bka,
	Pages = {751-761},
	Title = {Multiplicative censoring, renewal processes, deconvolution and decreasing density: nonparametric estimation},
	Volume = {76},
	Year = {1989}}

@article{warkany1978terathansaia,
  title={Terathansaia},
  author={Warkany, Josef},
  journal={Teratology},
  volume={17},
  pages={187-192},
  year={1978}
}

@article{wu:etal,
	Author = {Wu, Y and Chambers, C and Xu, R},
	Journal = lida,
	Pages = {507-528},
	Title = {Semiparametric sieve maximum likelihood estimation under cure model with partly interval censored and left truncated data for application to spontaneous abortion},
	Volume = 25,
	Year = 2019}

@article{xu:cham,
	Author = {Xu, R and Chambers, C},
	Journal = {Reproductive Toxicology},
	Pages = {490-493},
	Title = {A sample size calculation for spontaneous abortion in observational studies},
	Volume = 32,
	Year = 2011}

@article{xu2018impact,
  title={The impact of confounder selection in propensity scores when applied to prospective cohort studies in pregnancy},
  author={Xu, Ronghui and Hou, Jue and Chambers, Christina D},
  journal={Reproductive Toxicology},
  volume={78},
  pages={75--80},
  year={2018},
  publisher={Elsevier}
}

@article{xu2019statistical,
  title={Statistical sensitivity analysis for the estimation of fetal alcohol spectrum disorders prevalence},
  author={Xu, Ronghui and Honerkamp-Smith, Gordon and Chambers, Christina D},
  journal={Reproductive Toxicology},
  volume={86},
  pages={62--67},
  year={2019},
  publisher={Elsevier}
}

@article{yang2016using,
	Author = {Yang, Fan and Small, Dylan S},
	Journal = {Journal of the Royal Statistical Society: Series B (Statistical Methodology)},
	Number = {1},
	Pages = {299--318},
	Publisher = {Wiley Online Library},
	Title = {Using post-outcome measurement information in censoring-by-death problems},
	Volume = {78},
	Year = {2016}}

@phdthesis{ying2020statistical,
  title={Statistical Inference: Global Testing, Multiple Testing and Causal Inference in Survival Analysis},
  author={Ying, Andrew},
  year={2020},
  school={University of California, San Diego}
}

\end{document}